\documentclass{jfm}
\usepackage{graphicx}
\usepackage{epstopdf}
\usepackage{bm}
\usepackage{natbib}
\usepackage{amsmath}
\usepackage{color}
\usepackage[utf8]{inputenc}
\usepackage{dirtytalk}
\usepackage{float}
\usepackage{afterpage}

\def\hh{{\hat{h}}}

\def\half{{\mbox{$\frac12$}}}
\def\JJ{{\mathcal{J}}}
\def\BB{{\mathcal{B}}}
\def\RR{{\mathcal{R}}}

\begin{document}

\newtheorem{lemma}{Lemma}
\newtheorem{corollary}{Corollary}

\shorttitle{Viscoplastic Spreading} 
\shortauthor{M. Jalaal et al.} 

\title{Spreading of viscoplastic droplets} 

\author
 { Maziyar Jalaal\aff{1,2}
  \corresp{\email{mazi@alumni.ubc.ca, mj547@cam.ac.uk}},
Boris Stoeber\aff{2,3}
Neil J. Balmforth\aff{4},
  }  

\affiliation
{

\aff{1}
Department of Applied Mathematics and Theoretical Physics,
University of Cambridge, Cambridge, CB3 0WA, United Kingdom

\aff{2}
Department of Mechanical Engineering, University of British Columbia, BC, Canada

\aff{3}
Department of Electrical and Computer Engineering, University of British Columbia, BC, Canada

\aff{4}
Department of Mathematics, University of British Columbia, BC, Canada
}

\maketitle

\begin{abstract}
  The spreading under surface tension and gravity of a droplet
  of yield-stress fluid over a thin film of the same material
  is studied. The droplet
  converges to a final equilibrium shape once the driving stresses
  inside the droplet fall below the yield stress.
  Scaling laws are presented for the final radius and
  complemented with an asymptotic
  analysis for shallow droplets. Moreover,  numerical simulations using
  the volume-of-fluid method and a regularized constitutive law,
  and experiments with an aqueous solution of Carbopol are presented.
\end{abstract}

\section{Introduction}
\label{sec: intro}

{\color{black}
The impact and spreading of droplets of complex fluids over surfaces occur in a wide variety of industrial applications. Examples include, but are not limited to, inkjet printing, spray coatings, and additive manufacturing \citep{barnes1999yield, derby2010inkjet, mackay2018importance, thompson2014}. In many cases, such as 3D printers and paint sprays, liquid droplets or filaments are deposited on an existing layer of the same fluid. Hence, understanding the underlying fluid mechanics of droplet spreading on a thin film helps to improve the design of such systems.
On the theoretical side, the removal of an advancing contact
line has the extra advantage of simplifying
the spreading problem substantially by removing the complicated physics 
associated with relieving the stress singularity that otherwise arises
({\it e.g.} \cite{oron1997long, craster2009dynamics, bonn2009wetting}).
Precursor films are also expected to be
drawn out ahead of spreading droplets by intermolecular forces,
effectively emplacing a pre-wetted film even in situations in which
one did not exist originally.
({\it e.g.} \cite{craster2009dynamics, bonn2009wetting}).
}

Newtonian droplets spreading over a thin layer has been addressed previously for a range of physical regimes (see \cite{bergemann2018viscous} and \cite{jalaal2019capillary}) and the references therein). The present article aims to provide a discussion on the viscoplastic version of this problem. Viscoplastic
or yield stress fluids feature a mix of fluid and solid behavior:
if not sufficiently stressed, such materials behave more like an elastic material, but above a critical yield stress, they flow like a viscous fluid. Yield stress is a common feature of many natural and industrial fluids, such as clay, cement, toothpaste, cosmetic creams, dairy products, waxy oil, and many more (see Balmforth, Frigaard \& Ovarlez (2014), Coussot (2014) and Bonn et al. (2017) for reviews).\nocite{balmforth2014yielding,coussot,bonn2017}

For a Newtonian droplet deposited on a thin film of the same fluid
and spreading due to gravity and surface tension, flow 
continues until a completely flat film is formed. By contrast,
when the driving stresses fall below the yield stress,
flow ceases inside a viscoplastic droplet.
Consequently, the final shape is not flat, as shown previously in different configurations \citep{roussel2005fifty, balmforth2007viscoplastic1, german2009, chen2017morphology, jalaal2015slip,liu2016two, liu2018axisymmetric}.
The final shape of the droplets is of particular importance in industrial
processes such as 3D printing as it can determine the resolution of the printing process and the final quality of the product. Previous experiments have been reported for the problem of viscoplastic droplet impact on solid
surfaces \citep{luu2009, saidi2010influence, luu2013giant, blackwell2015sticking, sen2020viscoplastic} or fluid interfaces \citep{jalaal2019vpentry}.


In the present work, we explore the spreading of a viscoplastic droplet and its final shape, providing a theoretical framework for the problem complemented
with experimental and computational results.
The order of material in this paper is as follows. Section \ref{sec:scalinglaws} presents simple scaling laws for the final shape of a viscoplastic droplet. Section \ref{sec:VPLub} summarizes a viscoplastic lubrication theory suitable for
shallow droplets. Section \ref{sec:Num} presents the numerical simulations for a spreading viscoplastic droplet. Section \ref{sec:Exp} describes the experimental tests, and is followed by section \ref{sec:comp} which summarizes the theoretical and experimental results.
The Appendices contain further technical details of the lubrication
theory and numerical computations.


\section{Scaling Laws for the Final Shape}
\label{sec:scalinglaws}

Consider a viscoplastic droplet deposited  on a thin film of thickness
$H_{\infty}$ at $t=0$.
{\color{black} We imagine that the
  droplet yields entirely under 
  capillary action or gravity, spreading and then braking to a halt
  due to the yield stress $\tau_0$. That is, we consider the
  situation in which the yield stress cannot localized flow
  and leave intact a substantial volume of the droplet
  to imprint a dependence of the final shape on the initial
  configuration. Global force balance over the entire droplet volume
  should then control the
  final radius, $\mathcal{R}_f$, and height, $\mathcal{H}_f$.
If the rheology of the fluid only features through
the yield stress, the physical parameters of the problem}
include $\tau_0$,
the density of the droplet $\rho$,
the surface tension coefficient $\sigma$, gravitational acceleration $g$,
and the droplet volume $\mathcal{V}$.

%

When the droplet spreads under capillary effects, the driving
horizontal pressure force (given by the product of the
pressure $p\sim\sigma\kappa$ and a
typical vertical surface area $\mathcal{H}_f\mathcal{R}_f$)
can be estimated as
$$
p\mathcal{H}_f\mathcal{R}_f \sim \sigma \kappa \mathcal{H}_f\mathcal{R}_f
\sim \sigma \frac{\mathcal{H}_f^2}{\mathcal{R}_f},
$$
where  $\kappa \sim \mathcal{H}_f /  \mathcal{R}_f^2$,
is the curvature. On the other hand, if the droplet does not slip
over the underlying surface and
the yield stress acting over the base of the droplet
provides the main resistance to flow, the opposing force is of order
$\tau_0 \mathcal{R}^2_f$. By balancing the two, we arrive at
$$
\sigma \mathcal{H}_f^2 \sim \tau_0 \mathcal{R}^3_f.
$$
Moreover, since $\mathcal{V}\sim\mathcal{H}_f\mathcal{R}^2_f$,
if we define the lengthscale
$\mathcal{L}=[3\mathcal{V}/(4\pi)]^{1/3}$
({\it i.e.} the radius of the corresponding spherical drop),
then
\begin{equation}
  \sigma \mathcal{V}^2 \sim \tau_0 \mathcal{R}^7_f ,
  \quad {\rm or} \quad
  \frac{\mathcal{R}_f }{\mathcal{L}} =
  \Omega_c \, \mathcal{J}^{-1/7},
  \quad \mathrm{where} \quad \mathcal{J}=\frac{\tau_0 \, \mathcal{L}}{\sigma}  
\label{eqn:J}
\end{equation}
is a non-dimensional number that compares the yield stress and
capillary pressure. Thus,
as the plastic effect $\mathcal{J}$ increases,
the final radius becomes correspondingly smaller.
In (\ref{eqn:J}), the prefactor $\Omega_c$ encapsulates
dependence on the remaining
dimensionless groups in the problem.
If gravity and any other effects are not important, the only remaining group
is the scaled pre-wetted film thickness, $h_\infty \equiv H_\infty/\mathcal{L}$.

If we instead counter
a driving hydrostatic pressure $p\sim \rho \, g \, \mathcal{H}_f$
by the resistance from the yield stress,
the balance is
\begin{equation}
  \rho g \mathcal{H}_f^2 \mathcal{R}_f \sim \tau_0 \mathcal{R}^2_f,
  \quad {\rm or} \quad
  \frac{\mathcal{R}_f }{ \mathcal{L}}
  = \Omega_g \, \left( \frac{\rho g \mathcal{L}}{\tau_0} \right)^{1/5}
  \equiv \Omega_g \, \mathcal{B}^{1/5} \, \mathcal{J}^{-1/5} , \quad
  \mathrm{where} \quad  \mathcal{B} = \frac{\rho \, g \, \mathcal{L}^2}{\sigma}
\label{eqn:Bo}
\end{equation}
is the Bond number,
comparing the hydrostatic pressure and capillary pressure.
Again the prefactor $\Omega_g$ contains the dependence on any
other dimensionless groups.
Note that the combination $\mathcal{B} / \mathcal{J}$ is independent
of $\sigma$, eliminating surface tension from the right-hand side
of (\ref{eqn:Bo}) when capillary effects are not present and
$\Omega_g$ depends only on $h_\infty$.
This limit is relevant to geophysical flows and rheometry
with larger spatial scales
({\it cf.} \cite{revslump} and \cite{roussel2005fifty}). 

More generally, we must take either $\Omega_c$ or $\Omega_g$
to depend on both $\mathcal{B}$ and $h_\infty=H_\infty/\mathcal{L}$,
as well as any other dimensionless parameters stemming from
further physical effects.
In what follows, we assume that only $\mathcal{B}$ and $h_\infty$
are relevant and write the general relation,
\begin{equation}
  \frac{\mathcal{R}_f }{\mathcal{L}} =
  \Omega\left(\mathcal{B},h_\infty\right)
  \, \mathcal{J}^{-1/7} ,
\label{eqn:Ja}
\end{equation}
with $\Omega\to\Omega_c$ in the capillary-dominated limit $\mathcal{B}\to0$,
and $\Omega\to\Omega_g \, \mathcal{B}^{1/5}\mathcal{J}^{-2/35}$
in the gravity-dominated limit $\mathcal{B}\gg1$. In particular,
we compare this scaling law with
asymptotic analysis, numerical simulations and experiments,
each of which provide more refined estimates for $\Omega$.

\section{Viscoplastic Lubrication Theory}
\label{sec:VPLub}

\begin{figure}
    \centering
    \includegraphics[width=0.98\linewidth]{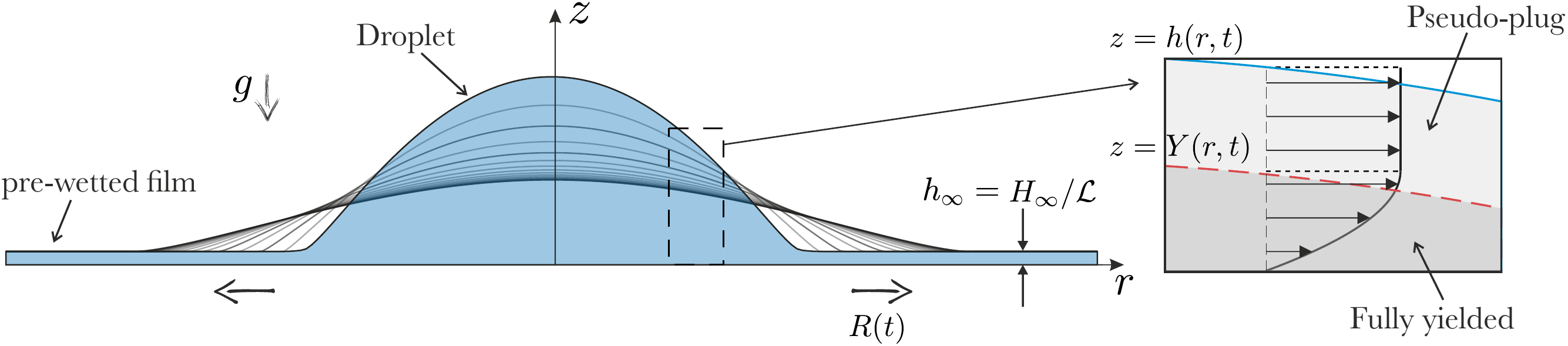}
    \caption{Sketches ot the geometry and anatomy of a
      spreading viscoplastic droplet.
      The top surface is $z=h(r,t)$; below $z=Y(r,t)$ the fluid
      is fully yielded and flows in a plug-like manner over
      $Y<z<h$ where fluid is held near the yield stress (the ``pseudo-plug'').}
    \label{fgr:LubSchem}
\end{figure}

As sketched in figure \ref{fgr:LubSchem}
and assuming that the droplet remains axisymmetrical,
we employ cylindrical polar coordinates $(r,z)$ to describe the geometry
of a shallow viscoplastic droplet.
The top surface of the fluid lies at $z=h(r,t)$, and the droplet
is emplaced upon an existing fluid layer of thickness $H_\infty$.
There is a rigid plane at $z=0$ over which the fluid cannot slip.
To simplify our discussion,
we used the Bingham constitutive law, which combines the
yield stress $\tau_0$ with a plastic viscosity $\mu$.

{\color{black}
Lubrication theory applies when droplets
are relatively shallow and inertia is negligible.
In this instance, the hydrostatic and capillary pressures
are largely independent of $z$ and drive spreading, which is
primarily countered by the vertical shear stress.
The analysis for viscoplastic fluid follows along similar lines
to that for Newtonian fluid
({\it e.g.} \citep{oron1997long, craster2009dynamics}), the
key differences arising from the impact of the yield stress
on the vertical profile of the radial velocity
\citep{liu1989slow,revslump}. In particular,
as illustrated in figure \ref{fgr:LubSchem}, 
the velocity field adopts a distinctive anatomy 
in which a lower slice of the fluid
in $0<z<Y(r,t)$ is fully yielded and the radial velocity
has a parabolic profile; over the region 
$Y(r,t)<z<h(r,t)$, the radial velocity becomes plug-like and independent
of $z$ to leading order. 
The upper region possesses a shear stress that lies below the yield stress,
an observation that has previously led to the incorrect conclusion
that the fluid here is unyielded, contradicting the radial expansion
and appearing to imply a paradox in lubrication theory.
In fact, over the plug-like region, or ``pseudo-plug'',
the extensional stresses are of the same order as the shear stress
(a feature demanded by the proper three-dimensional form
of the Bingham constitutive law when deformation rates are relatively small)
and their conspiracy holds the stress slightly
above $\tau_0$ to permit the radial expansion
\citep{balmforth1999consistent, putz2009lubrication}.
In the limit $\tau_0\to0$, the pseudo-plug disappears ($Y\to h$)
and we recover the fully parabolic Newtonian flow profile.
Conversely, when $Y\to0$, the pseudo-plug reaches the base,
bringing the entire fluid layer to a halt.
}

Given the shallow-layer velocity field of the lubrication analysis,
the expression of depth-integrated mass conservation provides an
evolution equation for the local fluid depth.
We express this equation in the dimensionless form,
\begin{equation}
 h_t = \frac{1}{6 r} \left[r\, p_r \,  Y^{2} \left( 3h-Y \right) \right]_r,
  \qquad p  =  \BB  \, h - \frac{1}{r}\left(r h_r \right)_r.
\label{eqn:thin}
\end{equation}
after scaling lengths by $\mathcal{L}$,
pressure $p(r,t)$ by $\sigma / \mathcal{L}$,
velocity by $U = \sigma/\mu$ and time by $\mathcal{L}/U$;
the  surface bordering the pseudo-plug is given by
\begin{equation}
  Y= \mathrm{max} \left( 0, h- \frac{\JJ}{ \vert p_{r} \vert} \right).
\label{eqn:Y}
\end{equation}
{\color{black}
For $\JJ=0$, $Y=h$ and (\ref{eqn:thin}) reduces to the standard 
evolution equation
for a Newtonian film 
and therefore recovers known spreading laws for viscous droplets
under the action of gravity, capillarity or a combination of both
\citep{oron1997long, craster2009dynamics,bonn2009wetting}.
}

\subsection{Sample spreading solutions}

To provide sample solutions for spreading viscoplastic droplets,
we solve (\ref{eqn:thin}) numerically using centred finite differences
to approximate radial derivatives
and a stiff integrator to step the surface profile forwards in time.
The length of the domain is chosen
sufficiently large that the droplet never reaches the border.
We begin from the initial condition
$h(r,0) = \mathrm{max} \left(  0,  1-  3r^2 /16 \right)^3 + h_{\infty}$,
which smoothly interpolates between
\textcolor{black}{
the pre-wetted film of scaled thickness $h_\infty$ and an initial ``bump''
  with a dimensionless radius $R(0)=4/\sqrt{3}$
  (chosen so that the dimensional volume is ${\cal V}=4\pi {\cal L}^3/3$,
  given the scaling of lengths by $\cal L$);
  the results
  are insensitive to the precise initial shape of the bump
  provided the droplet
  becomes fully yielded during spreading}.
We also replace (\ref{eqn:Y}) by the regularization,
$Y=$max($\varepsilon,h-\JJ/|p_r|)$, to ease computations,
with $\varepsilon=10^{-6}$ (having verified that the precise value
makes no significant difference to the results).

\begin{figure}
    \centering
    \includegraphics[width=0.95\textwidth]{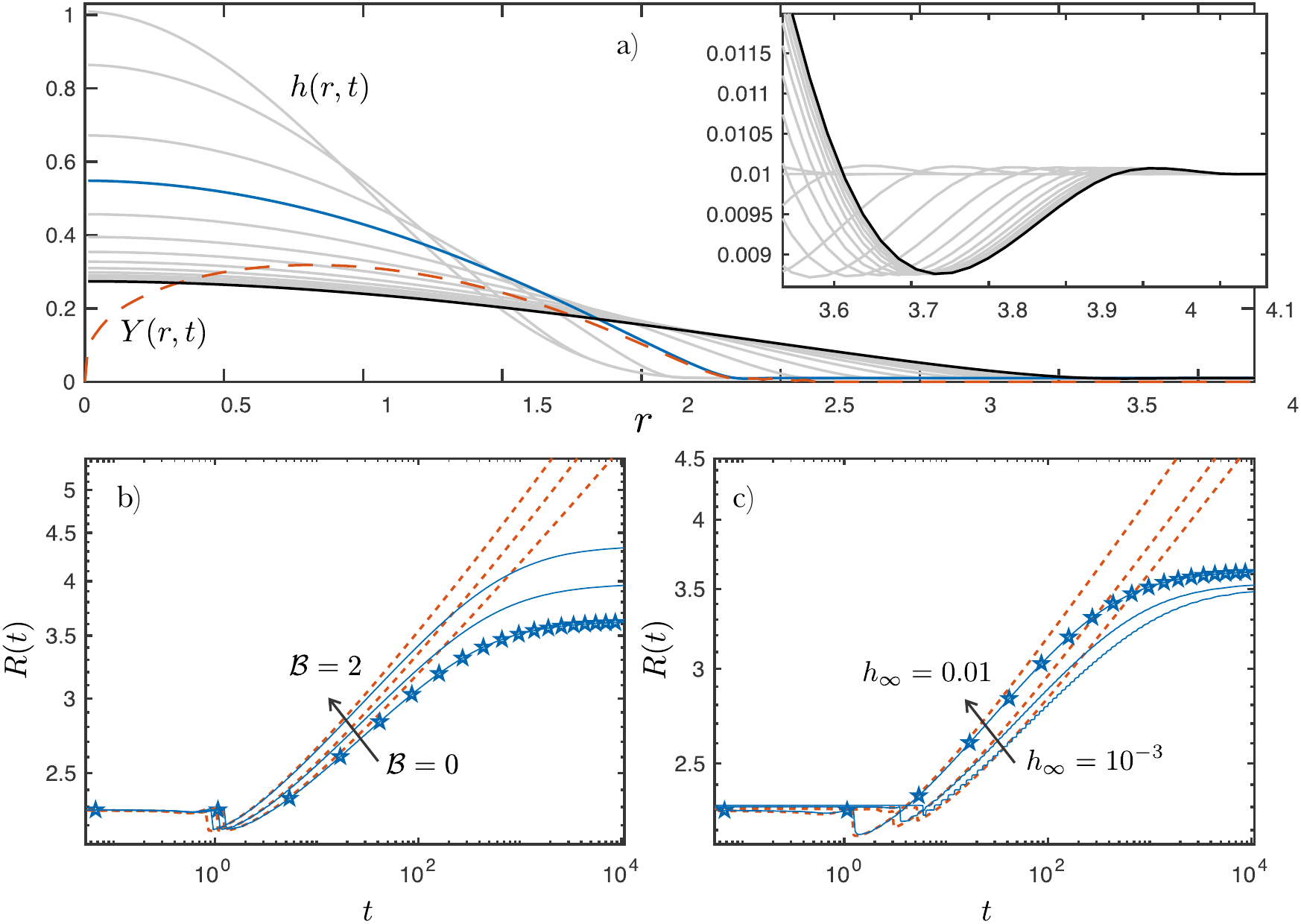}
    \caption{Numerical solutions of the evolution equation (\ref{eqn:thin}).
      (a) Snapshots of $h(r,t)$ (grey lines) for $\JJ=5/4$ and $\BB=0$,
      with the inset showing a magnification of the
      edge of the droplet; the final snapshot is plotted black. The
      surface $z=Y(r,t)$ at an intermediate time (corresponding
      to the curve of $h(r,t)$ in blue) is included
      as the (red) dashed line.
      Time series of the edge (defined as the first radial position $R(t)$
      where $h(R,t)=h_\infty$) for (b) $(\JJ,h_\infty)=(1.25,0.01)$
      with $\BB=0$, 3/4 and 2, and (c) $(\JJ,\BB)=(1.25,0)$ with
      $h_\infty=10^{-3}$, $3\times10^{-3}$ and $0.01$.
      The (red) dotted lines show corresponding Newtonian solutions ($\mathcal{J}=0$), and
      the stars indicate the times of the snapshots in (a).
      The relatively
      sharp features in $R(t)$ for $t\approx1$ in (b)-(c) correspond to the
      creation of the undulating wavetrain at the edge of the droplet.
    }
    \label{fgr:LubSim}
\end{figure}

Figure \ref{fgr:LubSim}(a) shows a solution
with $\JJ=5/4$ and $\BB=0$, using an initial fluid film of thickness
$h_\infty=0.01$. The dimensionless yield stress is sufficiently small
that the entire droplet yields under capillary action at $t=0$
and then spreads much like a Newtonian droplet. Subsequently,
however, the yield stress comes into play as driving stresses decline,
and eventually the droplet brakes to rest. 
Distinctive spatial oscillations appear near the edge of the droplet,
becoming frozen into the final shape as flow ceases. These
undulations also appear in the Newtonian problem and have been
reported previously for viscoplastic films
\citep{balmforth2007surface,jalaal2016long}. We discuss them further
below and in greater detail in appendix \ref{app:Edge}.

The solutions are much the same, though wider and flatter, with
gravity ($\BB>0$). Figure \ref{fgr:LubSim}(b) and (c) show
time series of the edge $R(t)$ for further solutions
with different parameter settings. Here, the edge is
measured for numerical convenience 
as the first location $R(t)$ where $h(R,t)=h_\infty$.
{\color{black} With $\JJ=0$, the edge continues to expand;
the yield stress, however, inevitably brings fluids to rest.}

\subsection{Final shapes}
\label{sec:asymptots}

A viscoplastic droplet comes to rest
when $Y \rightarrow 0$, implying $h |p_r| = \JJ$, which comprises a
third-order differential equation for the final profile
in view of (\ref{eqn:thin}).
A complication in solving this equation
is the presence of the factor $|p_r|$, which leaves open
the sign of the pressure gradient. To determine this sign,
we appeal to the evolution equation (\ref{eqn:thin}) and the limit of its
solution as the fluid comes to rest. In particular,
over the bulk of the droplet,
the sign of $-p_r$ corresponds to the sense of the flux, which must be
positive. Near the edge, however, the spatial oscillations complicate matters.
There, as is clear from the magnification
in figure \ref{fgr:LubSim}(a), the undulations corresponds to a travelling
wavetrain such that
\begin{equation}
h_t \to  - R_t h_r  \sim \frac{1}{6} [ p_r Y^2 (3h-Y)]_r .
\end{equation}
Integrating this equation and observing that $h\to h_\infty$
for $p_r Y^2 (3h-Y)\to0$, we find that the sign of $-p_r$ must be
given by the sign of $h-h_\infty$.
Thus,
\begin{equation}
  h_{rrr} + \frac{1}{r} h_{rr} - \frac{1}{r^2} h_r
  - \BB  \; h_r 
  = \frac{\JJ}{h}\; {\rm sgn}(h-h_\infty) .
\label{eq:finalshapeEq}
\end{equation}

\subsubsection{Gravity-dominated limit}
\label{3.2.1}

In the gravity-dominated limit, the higher derivatives disappear from
the left-hand side of
(\ref{eq:finalshapeEq}), leaving $- hh_r \sim \JJ/\BB$ (since
$h\geq h_\infty$). Thence
\begin{equation}
  h =  \left[ h_\infty^2 +  \frac{2\JJ}{\BB} (R-r)  \right]^{1/2}
\label{eqn:gravityfinal} 
\end{equation}
({\it cf.} \cite{blake1990viscoplastic, roussel2005fifty, revslump}).
The corresponding droplet volume must be
$\mathcal{V}=2 \pi \mathcal{L}^3 \int_0^R [h(r)-h_\infty] r\;{\rm d}r$,
which demands that the final radius satisfy
the algebraic problem,
\begin{equation}
  R\equiv\frac{\mathcal{R}_f}{\mathcal{L}} =
  \left( \frac{25\, \BB}{8 \, \JJ}\right)^{1/5}
  \left[ (1+A)^{5/2} - \half A^{3/2} (2A+5)
    - \mbox{$\frac{15}{8}$}A^{1/2}\right]^{-2/5},
  \qquad  A = \frac{\BB \, h_\infty^2}{2\, \JJ R} .
\end{equation}
The solution can be written formally as
$\RR_f/\mathcal{L} =
\Omega_g\left(h_\infty\sqrt{\BB/\JJ}\right) \BB^{1/5}\JJ^{-1/5}$
in the manner of (\ref{eqn:Bo}). Notably, when $h_\infty\to0$,
$\Omega_g\to(25/8)^{1/5}\approx1.26$.

\subsubsection{Finite $\BB$}
\label{3.2.2}

Away from the gravity-dominated limit, we must attack (\ref{eq:finalshapeEq})
numerically. For a pre-wetted film with finite thickness, the boundary
conditions are the symmetry condition $h_r(0)=0$
and the far-field condition, $h\to h_\infty$ for large radii.
The latter requires the imposition of two conditions
in order that the droplet profile meets the pre-wetted film continuously.
Given the form of the solution at the edge,
we choose $h(R_*)=h_\infty$ and $h_{rr}(R_*)=0$, where $R_*$
denotes a radius well into the decaying undulations. Applying this
boundary condition corresponds to pinning
the solution at a point further along the wavetrain.
In addition, we must also arrive at the correct droplet volume.
Thus, the third-order equation (\ref{eq:finalshapeEq})
must be solved subject to three boundary conditions and the volume
constraint, demanding that the edge position $R_*$ be found
as part of the solution ({\it i.e.} an eigenvalue).
Appendix \ref{app:Edge} describes further details of the
numerical construction of the final profile, as well as a more detailed
consideration of the undulations at the edge.

\begin{figure}
    \centering
    \includegraphics[width=0.95\textwidth]{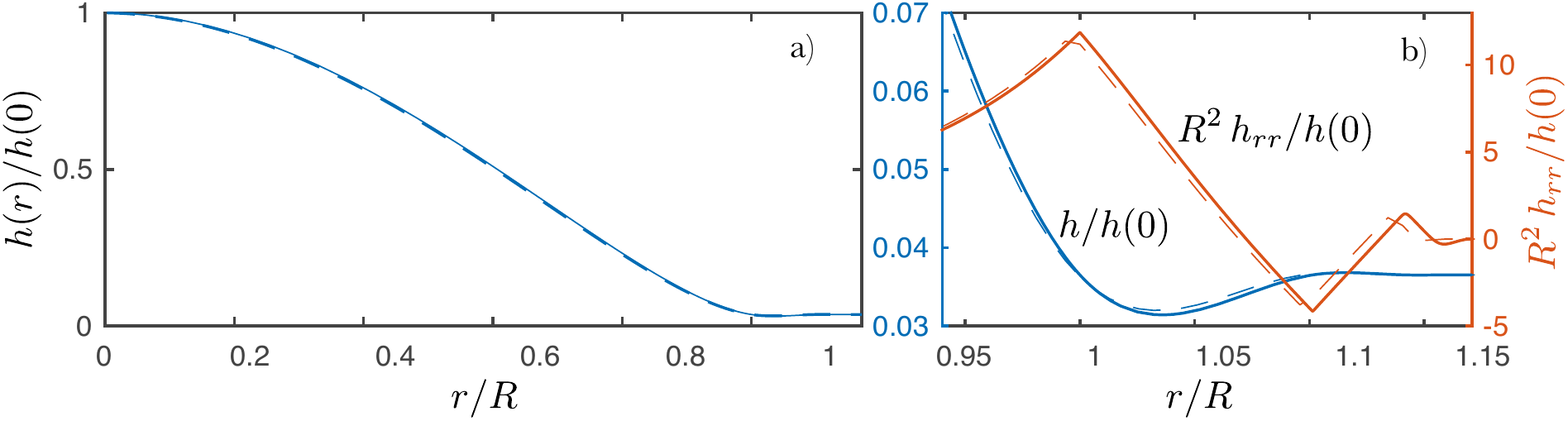}
    \caption{
      (a) A final equilibrium profile for $(\JJ,\BB,h_\infty)=(1.25,0,0.01)$.
      A magnification near the edge is shown in (b) along with
      $h_{rr}$. The dashed lines show the final snapshot of the
      numerical solution of the evolution equation from
      figure \ref{fgr:LubSim}(a).
      }
    \label{exoplot}
\end{figure}

Figure \ref{exoplot} shows a
 sample numerical solution with $\BB=0$.
This particular example corresponds to the solution of the evolution
equation in figure \ref{fgr:LubSim}(a), and is compared with
the final snapshot of that computation in figure \ref{exoplot}.
The decaying undulations converge to a sawtooth wave in $h_{rr}$,
with the corners corresponding to the sign switches of $h-h_\infty$.
Unlike the decaying capillary waves of moving Newtonian contact lines
which have fixed wavelength
\citep{tanner1979spreading, tuck1990numerical, jalaal2019capillary},
the viscoplastic undulations shorten with distance along the wavetrain.
In Appendix \ref{app:Edge}, we outline how the waveform
converges to piecewise cubic polynomials, with an accumulation point at a finite
outer radius.

\begin{figure}
    \centering
    \includegraphics[width=0.9\textwidth]{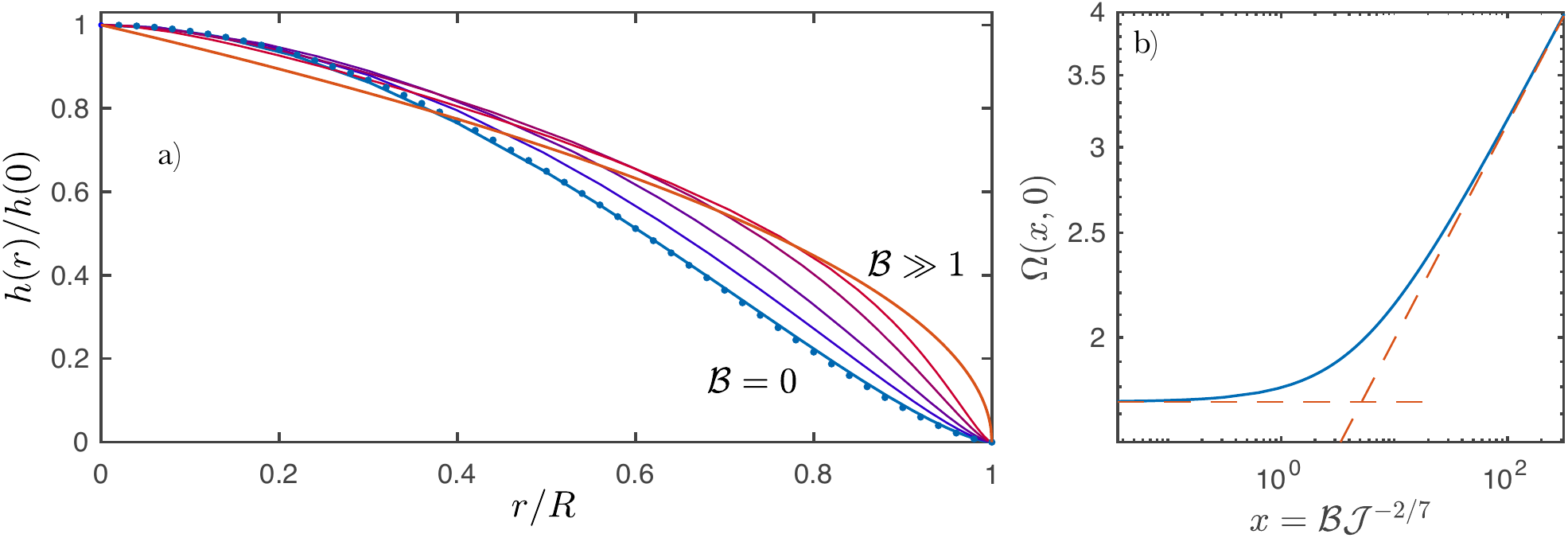}
    \caption{
      (a) Equilibrium profiles for $h_\infty\to0$ with
      $R^2\BB=0$, 10, 30, 100 and 300, together with the limit
      $\BB\to\infty$ given by (\ref{eqn:gravityfinal}).
      The dots show the approximation $(1-r^2)^{3/2}$.
      (b) The coefficient $\Omega(x=\BB\JJ^{-2/7},0)$ in
      (\ref{impo}), along with the limits for $x=0$ and $x\gg1$.      }
    \label{sossoplot}
\end{figure}

The analysis of the edge behaviour in Appendix \ref{app:Edge}
also indicates that the wavetrain shrinks to a point in the
limit $h_\infty\to0$. Moreover, the limiting solution
corresponds to solving (\ref{eq:finalshapeEq})
with outer boundary conditions based on the local solution
$h \sim \mathcal{C}(R-r)^{3/2}$ for $r\to R$ and an unknown
constant $\mathcal{C}$.
Figure \ref{sossoplot} illustrates such limiting profiles for
various values of the gravity parameter.
Evidently, since $h_r(R)=0$, the limiting contact angle is zero here,
implying spreading over a perfectly wetting surface
(although one can also solve (\ref{eq:finalshapeEq})
with a prescribed contact angle).

\def\LL{{\mathcal{L}}}
\def\OO{{\mathcal{F}}}
\def\HH{{\mathcal{H}}}

\begin{figure}
    \centering
    \includegraphics[width=0.8\textwidth]{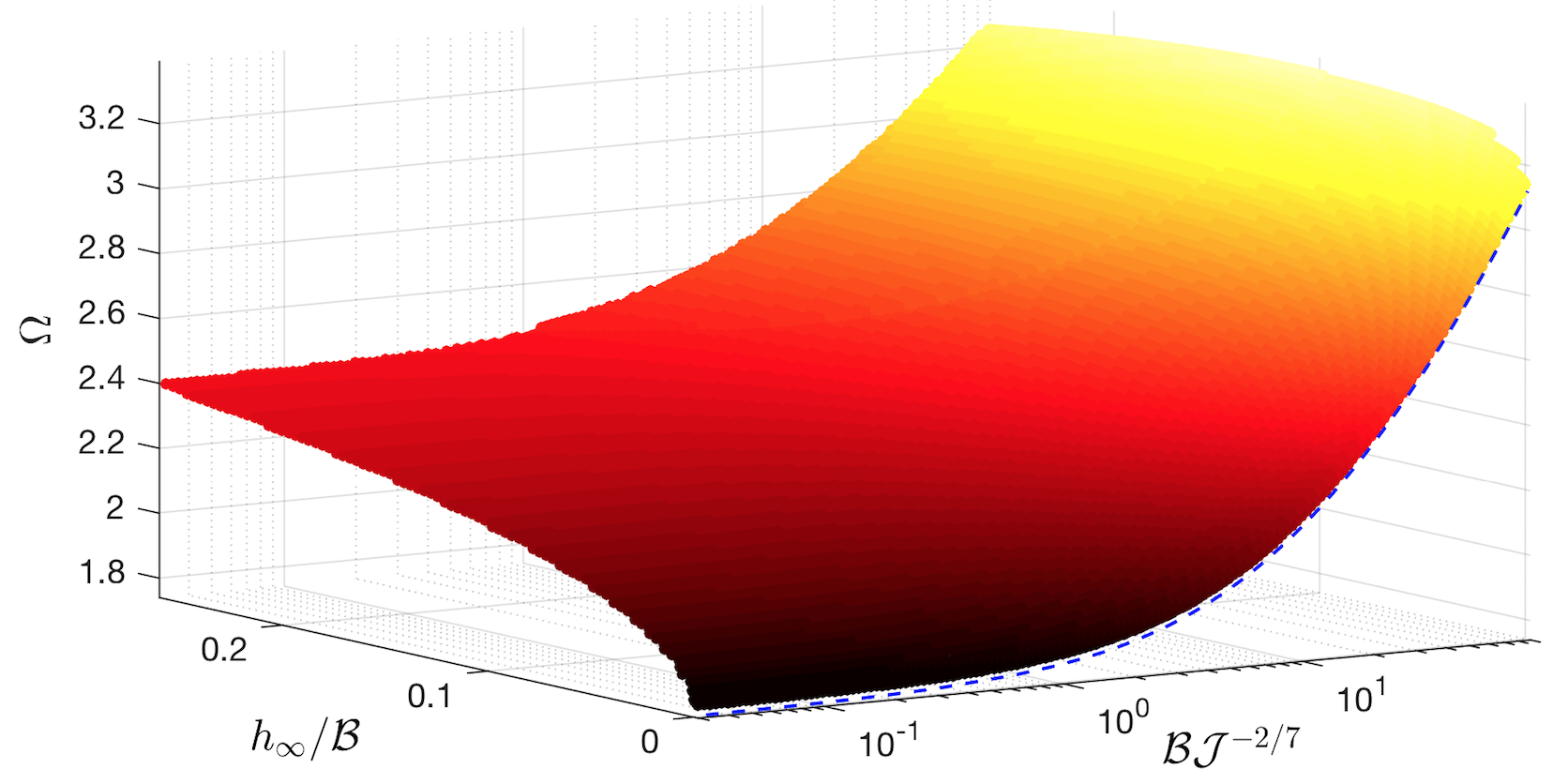}
    \caption{
      The function $\Omega({\BB}{\JJ^{-2/7}},{h_\infty}/{\BB})$ in (\ref{impo})
      plotted as a surface over the
      $({\BB}{\JJ^{-2/7}},{h_\infty}/{\BB})-$plane.
      The dashed line shows the result from figure \ref{sossoplot}(b).}
    \label{daplot}
\end{figure}

From the computed solutions for final equilibrium profiles, we may evaluate
the final radius and depth, which
are given by the (fairly complicated) algebraic problem
outlined in Appendix \ref{app:Edge}. For the final radius,
the formal solution may be written as
\begin{equation}
  \frac{\mathcal{R}_f}{\mathcal{L}} =
  \Omega\left(\frac{\BB}{\JJ^{2/7}},\frac{h_\infty}{\BB}\right) \JJ^{-1/7}
  ,
\label{impo}
\end{equation}
which identifies the dependence of the coefficient on gravity
and the prewetted film thickness.
The pre-factor $\Omega({\BB}{\JJ^{-2/7}},{h_\infty}/{\BB})$
is shown as a surface over the $({\BB}{\JJ^{-2/7}},{h_\infty}/{\BB})-$plane
in figure \ref{daplot}. In the limit $h_\infty\to0$, 
the function $\Omega(x,0)$ is shown in more detail
in figure \ref{sossoplot}(b); $\Omega_c\equiv\Omega(0,0)\approx 1.74$
and $\Omega(x,0)\to (25x/8)^{1/5}$ for $x\gg1$, which aligns
with the result in \S \ref{3.2.1}.

\section{Numerical Simulations}
\label{sec:Num}

To complement the lubrication analysis, we
solve the spreading problem numerically away from the shallow limit
using the open-source code Gerris \citep{popinet2003gerris}.
The code employs a Volume-of-Fluid (VOF) scheme to deal with the interface 
and an adaptive grid to achieve high resolution inside the droplet and
along its interface (see appendix \ref{app:VOF} for more details). 
For the rheology, we use the regularized Bingham model with
{\color{black}
\begin{equation}
  \tau_{ij} = \mu_1 \dot\gamma_{ij},
  \qquad
  \mu_1 = {\rm Min} \left( \frac{\tau_0}{\dot\gamma}
  + \mu , \; \mu_{max} \right) \dot\gamma_{ij}
,
\label{regul}
\end{equation}
where
\begin{equation}
  \dot\gamma_{ij} = \frac{\partial u_i}{\partial x_j} +
\frac{\partial u_j}{\partial x_i},\qquad
\dot\gamma\equiv \left( \half\sum_{i,j}\dot\gamma_{ij}\dot\gamma_{ji} \right)^{1/2}
\end{equation}
}
({\it cf.} \cite{o1984numerical}).
Here, the divergence of the viscosity at low shear rates is controlled by 
the regularization parameter, $\mu_{max}$, chosen sufficiently large
to render the simulations insensitive to the precise value
(see appendix \ref{app:VOF}). 
We use a parabolic initial shape for the droplet, merged to a flat pre-wetted
film, motivated by the experimental profiles observed in \S \ref{sec:Exp}:
$$
h(r,0)  = h_\infty +
(2/3)^{1/3} \, \mathrm{max} \left(0,1-(r/R_0)^2\right) ,
\quad \mathrm{with} \quad R_0=(2/3)^{1/3}.
$$
The simulation includes inertia, allowing us to take
the velocity field to be zero at $t=0$.  
However, for the physical parameters studied here, the effect of inertia is always small.

\begin{figure}
    \centering
    \includegraphics[width=370pt]{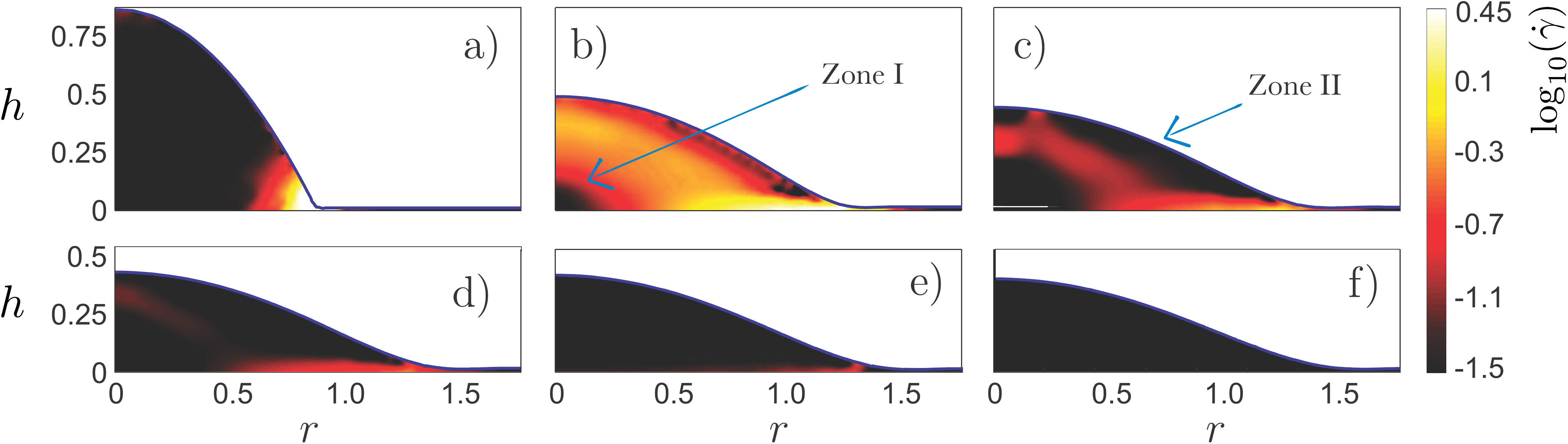}
    \caption{a-f) Distribution of $\mathrm{log}_{10}(\mathcal{\dot{\gamma}})$ in a
      spreading viscoplastic droplet for $\mathcal{J}=0.144$, $\mathcal{B}=0$
      and $h_\infty=0.0175$, at the times $t=0.29$, 87, 290, 580, 1450, 2030.}
    \label{fgr:NUM}
\end{figure}

Figure \ref{fgr:NUM} shows an example for $\mathcal{J}=0.144$, $\BB=0$
and $h_\infty=0.0175$,
plotting snapshots of the interface with superposed density maps of $\dot{\gamma}$.
Initially, the surface tension arising due to the
high curvature at the edge of the droplet drives flow, flattening the
entire profile over later times. Throughout, a plug remains at the
core of the droplet (marked as zone I in figure \ref{fgr:NUM}b),
much like in the gravity-driven problems
explored by \citep{liu2016two, liu2018axisymmetric}.
Once the droplet becomes shallower,
a further region of low strain rates forms close to the interface
(denoted as zone II in figure \ref{fgr:NUM}c),
resembling the pseudo-plug region of the lubrication analysis.
As the droplet brakes to rest, the regions of small strain rate broaden to
span the droplet,
{\color{black} with a train of undulations appearing at the edge
  as in the lubrication analysis
  (see Appendix B and figure \ref{fgr:GerrisRegul}).}
We calculate final shapes for a range of $\mathcal{J}$ and $\mathcal{B}$,
and later in section \ref{sec:comp}, we compare them with those obtained
from the lubrication analysis and the experiments.

\section{Experiments}
\label{sec:Exp}

\begin{figure}
    \centering
    \includegraphics[width=350pt]{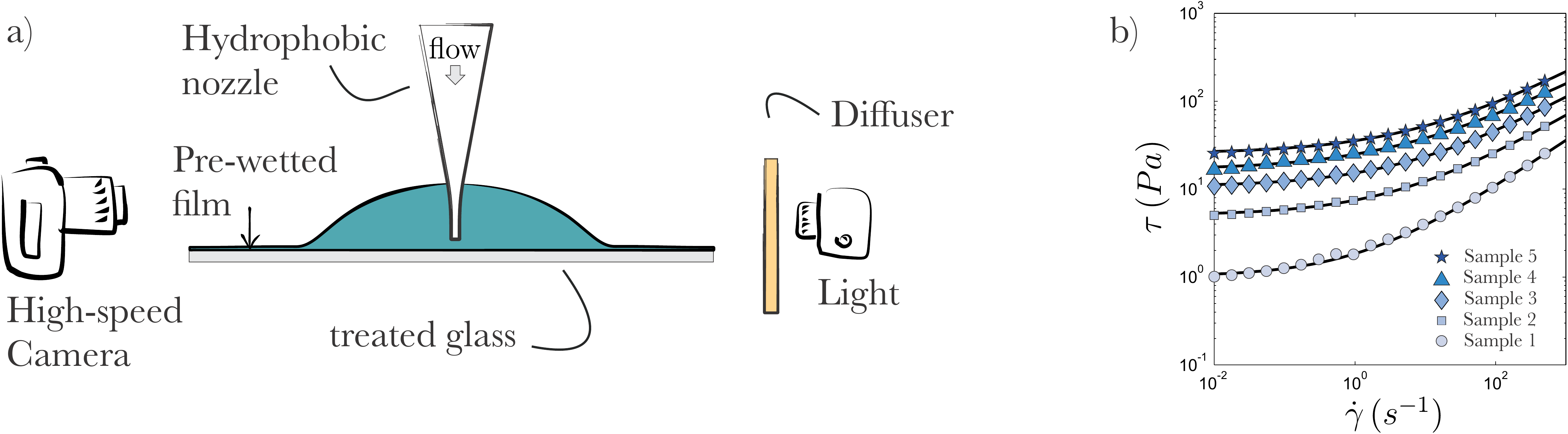}
    \caption{a) Schematic picture of the setup for the extrusion tests, where the nozzle tip is placed just above the substrate. b) Flow curves of the fluids used in the tests on a log-log scale. The black curves show the Herschel-Bulkley fits.}
    \label{fgr:Exp1}
\end{figure}

We experimentally study the spreading of viscoplastic droplets by extruding
them from a syringe onto a pre-wetted surface. 
The experimental apparatus consisted of a hydrophobic nozzle (with inner diameter of 0.3\,mm) connected to a syringe pump (KD Scientific-Legato 111), where the vertical location of the nozzle could be adjusted using a translation stage.
Chemically treated glass slides were used to suppress any slip
over the underlying surface (see \cite{jalaal2015slip} for details).
To form the pre-wetted film, spacers of a given height (adhesive tape of $\sim$60 $\mu$m thickness ) were placed on either side of the surface, and a
mound of the fluid was spread out evenly with a flat blade. 
The experimental setup is schematically illustrated in figure \ref{fgr:Exp1}a.

To minimize inertial effects, we slowly extruded the droplets on the surfaces. The nozzle tip was placed 200$\mu m$ above the film, and droplets of different volumes ($\mathcal{V}=$0.0042 to 1.4367 mL) were deposited. In all experiments, the extrusion flow rate was 2mL/min. The shapes of the droplets during spreading were recorded using side imaging, a cold LED light source illuminating the test section, and images were recorded with a high-speed camera attached to a microscope. The shape and volume were obtained through image processing.

\begin{table}
\label{carbopolrheo}
 \begin{center}
   \caption{Properties of the experimental fluids.}
  \begin{tabular}{ccccc}
    Sample number  & $n$  &  $K$ (Pa $\cdot$ s$^{n}$)  &  $\tau_{0}$ (Pa) & $G$ (Pa) \\[3pt]
       1   & 0.541 & 0.834  & 1.01  & 9$\pm$2\\
       2   & 0.437 & 2.455  & 5.03  & 33$\pm$2\\
       3   & 0.448 & 4.579  & 10.82 & 57$\pm$1\\
       4   & 0.410 & 8.354  & 16.60 & 64$\pm$1\\
       5   & 0.427 & 10.080 & 25.46 & 66$\pm$1\\
  \end{tabular}
    \label{tbl:materials}
 \end{center}
\end{table}

The working fluids were aqueous suspensions of Carbopol Ultrez 21
(by Lubrizol), neutralized with triethanolamine. The rheology
of the fluids was characterized using controlled shear-rate
tests with an Anton Paar (Physica MCR-302) rheometer fitted
with sand-blasted (PP25-S) parallel plates (roughness of $\sim$4 $\mu$m).
Figure \ref{fgr:Exp1}b shows the flow curves of the five concentrations
of Carbopol used, plus Herschel-Bulkley fits of the form
$\tau = \tau_0 + K\,\dot{\gamma}^n$ to the measured
shear stress $\tau$ and shear rate $\dot\gamma$. The fitting parameters,
(the yield stress, consistency index $K$, and the flow index $n$)
are provided in Table \ref{tbl:materials}, supplemented by
estimates of the elastic moduli, $G$, found from
constant, low shear-rate tests (monitoring the stress growth with strain).

The density of the solutions is very close to water. 
Measuring the surface tension of yield stress fluids is challenging. For carbopol solutions, reported values vary over a range of $\sim 51-69$ mN$/$m \citep{manglik2001dynamic, boujlel2013, geraud2014, jorgensen2015yield}. Here, we did not
measure the surface tension of our samples and used the fiducial value of $63$ mN$/$m as the rough average of all the reported values.

\begin{figure}
    \centering
    \includegraphics[width=350pt]{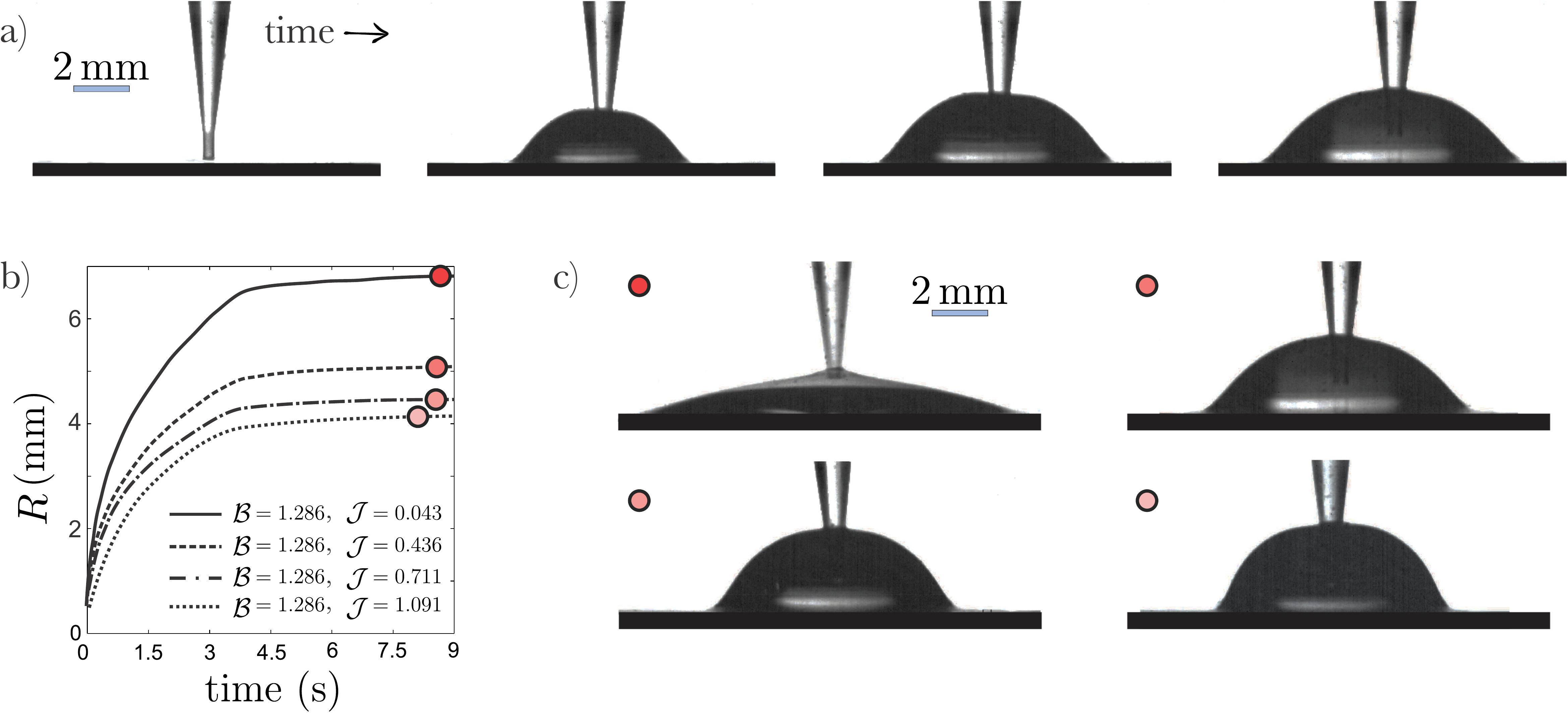}
    \caption{a) An example of an extrusion test for $\mathcal{B}=1.286$ and $\mathcal{J}=0.463$. Droplets start to grow with the pumping and reach a final shape quickly after the extrusion is over. Snapshots are at $t=0\,$s, $t=1.5\,$s, $t=3\,$s, and $t=4.5\,$s, respectively \textcolor{black}{(see video 1 in the supplementary material)}. The surface is pre-wetted with $h_{\infty} \sim 300 \mu$m.
b) The variations of the radius of droplets over time. Pumping is terminated at $\sim 3.5$s. The final state of the droplets are shown in panel c.}
    \label{fgr:Exp2}
\end{figure}

Figure \ref{fgr:Exp2}a displays an example extrusion test. A small droplet forms at the beginning of the injection and grows in time as the pumping continues. When the pump is switched off, the droplet keeps spreading freely until reaching the final state. Figures \ref{fgr:Exp2}b and c show results from tests for a given $\mathcal{B}$ at different $\mathcal{J}$ (same volume with different yield stress). As expected,
an increase in the magnitude of $\mathcal{J}$ results in a smaller final radius of the droplet.  Note that the photograph of the final shapes were taken
30\,s after the extrusion commenced, to ensure the equilibrium shape
was reached.
{\color{black} Although some success has been enjoyed with Newtonian fluid
  \citep{jalaal2019capillary}, we were also unable to clearly visualize
  any finer structure at the droplet edge such as the undulations
  that emerge in the lubrication theory and numerical simulations.}

Experiments were conducted for different $\mathcal{J}$ (by changing the
yield stress; {\color{black} {\it cf.} figure \ref{fgr:Exp1}b})
and $\mathcal{B}$ (by changing the droplet size), 
{\color{black} achieving 18 different parameter settings in total, as
  restricted by the limitations outlined below ({\color{black} {\it cf.} inset in figure \ref{fgr:res}}).}
The experiments were repeated (at least 5 times) for each data point, and the average values were calculated. The standard deviations (resulted from the accuracy of the image processing and repeatability of the tests) are small ($\sim 5-7$\%) for the majority of the extrusions. The standard deviation is however, larger, $\sim 15$\% at the lowest values of $\mathcal{B}$ since the drops are smaller and controlled deposition is harder. Note that the dimensional groups have limited ranges: to attain small values of $\mathcal{B}$, we had to extrude a small volume, promoting the effect of the nozzle. For larger values of $\mathcal{B}$, with our lowest yield stress, the droplets acquired a very large footprint that exceeded the boundary of the biggest chemically treated substrate available to us.

%

\section{Discussion and Conclusion}
\label{sec:comp}

\begin{figure}
    \centering
    \includegraphics[width=0.9\textwidth]{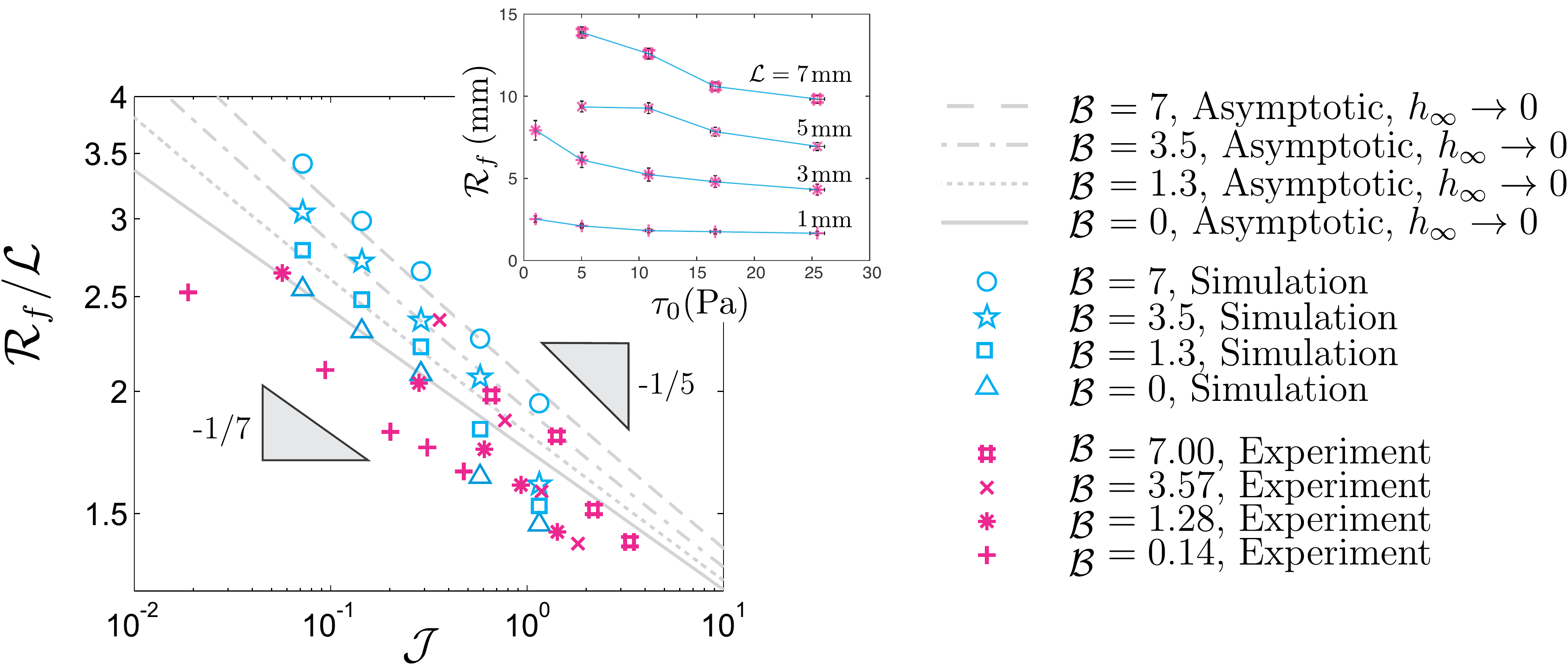}
    \caption{Comparison of the experimental and theoretical results of the final radius for different $\mathcal{J}$ and $\mathcal{B}$. Grey lines are asymptotic for $h_{\infty} \rightarrow 0$. \textcolor{black}{The inset shows the dimensional final radius for different yield stress and volume.}The results of $\mathcal{B}=0$ and $\mathcal{B}=0.14$ in asymptotic and simulations were so close so we just show the former for clarity.}
    \label{fgr:res}
\end{figure}

Figure \ref{fgr:res} summarizes our results for the dimensionless final radius
$\mathcal{R}_f/\mathcal{L}$ obtained from the lubrication theory, numerical
simulations and experiments. In the simulations, the value of the pre-wetted film is set at $h_{\infty}\approx 0.0175$.
In experiments, we measure this value to be $h_{\infty} \approx 0.07 \pm 0.03$. The results of the lubrication theory are shown for $h_\infty\to0$.
The three sets of results
show broad agreement with the predictions of scaling theory,
and, in particular, the trends $\JJ^{-1/7}$ and $\JJ^{-1/5}$ predicted
in the capillary and gravity dominated limits.

Although the theoretical and experimental results mostly overlap in
figure  \ref{fgr:res}, there are discrepancies. First, for the range of $\mathcal{B}$ explored here,
from figure \ref{daplot}, we see that the prefactor increases
by about 10\% if one increases the pre-wetted film thickness
from the limit $h_\infty\to0$ upto $h_\infty=0.0175$. Consequently,
the results from the lubrication theory should be about 10\% higher
in figure \ref{fgr:res} to match up properly against the simulations.
Evidently, the asymptotics overpredict the final radii,
a discrepancy that must originate in the shallow approximation of the lubrication analysis.

Besides this, the experimental data also fall somewhat below
the results from the simulations. This is surprising given that the
pre-wetted film thickness in the experiments is larger than
that used in the simulations, and a larger film thickness is expected to
furnish larger final radii. In other words, the experimental drops
definitely do not spread as far as suggested by the simulations.
This second discrepancy might have several origins. For instance, in
our simulations, we modelled a freely spreading droplet.
In the experiments, however, the droplets are extruded on the surface
from a syringe. The deposition method might therefore be responsible.
The difference is, in fact, more pronounced when the yield stress is
stronger (larger values of $\mathcal{J}$), as also notable in
figure \ref{fgr:res}. Indeed, in the simulations with the largest
yield stresses, not all of the droplet yields during spreading,
leaving intact significant plugged regions. By contrast, the action
of extruding the Carbopol through the syringe forces the fluid to
yield everywhere.

The fact that the flow history may impact the final state through
the evolution of the plugged regions
raises a second potential source for the discrepancy: the simulations exploit
the Bingham model whereas the Carbopol suspensions clearly have a nonlinear
plastic viscosity (see the Herschel-Bulkley fits in Table 1).
\textcolor{black}{The shear-thinning viscosity
may impact the final state by affecting the evolution of the plugs,
even if that rate-dependent effect is not expected to contribute
to the final force balance, or by affecting the dynamics at a nearly
singular contact line (\textit{cf.}
\cite{king2001a,king2001b, rafai2004spreading}).}
Worse, Carbopol is not an ideal yield-stress fluid. In particular,
this material has been
reported previously to be viscoelastic and sometimes thixotropic
\citep{balmforth2014yielding,coussot,bonn2017,dinkgreve2016different, coussot2002avalanche, luu2009,fraggedakis2016yielding},
all of which might contribute to the discrepancy in final radii.

In addition to the points above, there are some other technical difficulties in the experiments that might be responsible:  at the lowest $\mathcal{J}$, the
nozzle may have affected the spreading and final shape, especially when the droplet were small. Moreover, there are uncertainties in the experimental values
of the surface tension coefficient and yield stress that may affect
the dimensionless parameters. Additionally, these values are measured for a liquid bulk and it is not clear if they hold for small and confined geometries ({\it e.g.} \cite{geraud2013confined}).

To conclude,
in this paper, we have studied the spreading of a viscoplastic droplet on a thin film. In contrast to Newtonian droplets, a viscoplastic droplet spreading on a pre-wetted surface reaches a final shape, suggesting a practical means
to control the final radius by tuning the yield stress.
To gauge that final radius as a function of the physical
parameters of the problem,
we have provided simple scalings laws, a lubrication theory for
shallow droplets, numerical simulations for deeper droplets
and experiments with a Carbopol gel.
Our study has direct applications in many industries, such as 3D printing
and coating processes,
in which the spreading of droplets of yield-stress fluid plays a key role.
Possible extensions of our work include the development of
frameworks for more complicated constitutive models
such as elasto-viscoplastic models ({\it e.g.} \cite{saramito2007new, fraggedakis2016yielding, dimitriou2019canonical, oishi2019normal, oishi2019impact}),
and the use of more advanced experimental tools
to accurately measure droplet height and visualize the
internal flow field.

\section*{Acknowledgements}
MJ acknowledges the support of Natural Sciences and
Engineering Research Council of Canada through Vanier
Canada Graduate Scholarship.

\appendix

\section{Viscoplastic final shapes and contact lines}
\label{app:Edge}

\subsection{Computational details}

We may streamline the path to the solution of (\ref{eq:finalshapeEq})
by introducing the new variables $\xi=r/R_*$ and $\eta(\xi)=h(r)/h(0)$.
The task is then to solve
\begin{equation}
  \eta''' + \xi^{-1} \eta'' - \xi^{-2} \eta' - R_*^2 \BB \; \eta'
  = \frac{\Lambda_*}{\eta}{\rm sgn}\left( \eta - \hh_\infty \right)
  \label{A1}
\end{equation}
with eigenvalue and film thickness parameter,
\begin{equation}
  \Lambda_*\left(R_*^2 \BB,\hh_\infty\right) = \frac{R_*^3 \JJ}{[h(0)]^2} 
  \quad {\rm and} \quad
  \hh_\infty = \frac{h_\infty}{h(0)} ,
  \label{A2}
\end{equation}
and the boundary conditions,
\begin{equation}
  \eta(0)=1, \quad \eta'(0)=0, \quad \eta(1) = \hh_\infty ,\quad
  {\rm and} \quad \eta_{rr}(1)=0 .
\end{equation}
From this solution, we may then find $\eta_*$,
the first (scaled) radial position where $\eta(\xi_*)=\hh_\infty$,
corresponding to $r=R$.
A simple rescaling then provides $h/h(0)$ as a function of
$r/R$, from which we may compute the functions,
\begin{equation}
  \Lambda\left( R^2 \BB, \hh_\infty \right) =
  \Lambda_*\left( R_*^2 \BB, \hh_\infty \right)
  \quad {\rm and} \quad
  \mathcal{I}\left( R^2 \BB, \hh_\infty \right) =
 \frac{\int_0^R [h(r)-h_\infty] r\; {\rm d}r }{R^2 h(0)} .
\end{equation}

We use MATLAB's boundary-value-problem solver BVP4C to solve (\ref{A1}).
To further ease the computation we smooth out the
switch of sign on the right-hand side
using the function $\tanh[\Xi(\eta-\hh_\infty)]$, with $\Xi=10^6$,
which is sufficiently large to ensure the solution details are insensitive
to the precise value.
For the initial guess for the solver, we use either a final snapshot
from the solution of the evolution equation, or continuation
from a equilibrium profile with different parameter settings.
For the solution shown in figure \ref{exoplot}, the initial
guess is taken from the final snapshot in figure \ref{fgr:LubSim}(a),
and contains a sufficient number of undulations at the edge
such that the solver converges to an equilibrium profile with
five switches of sign of $h-h_\infty$. The slight smoothing of those
switches is visible in the plot of $h_{rr}$ at the right of panel (b).
To map out the functions, $\Lambda\left(R^2 \BB,\hh_\infty\right)$
and $\mathcal{I}\left(R^2 \BB,\hh_\infty\right)$, as required below,
we accelerate the computations by using a shorter
initial guess with only three sign switches.

\def\II{{\mathcal{I}}}

The final step is to consider the volume constraint,
which becomes
\begin{equation}
  \mathcal{V}= 2 \pi \RR_f^2\mathcal{H}_f \; \mathcal{I} (R^2 \BB, \hh_\infty).
\end{equation}
In conjunction with (\ref{A2}), we may then write
\begin{equation}
  \frac{\RR_f}{\LL} = \left( \frac{4\Lambda}{9\II^2}\right)^{1/7} \JJ^{-1/7}
    \quad {\rm and} \quad
  \frac{\HH_f}{\LL} = \left( \frac{8}{27\Lambda^2\II^3}\right)^{1/7} \JJ^{2/7}
  .
  \label{A6}
\end{equation}
The dependence of the functions $\Lambda$ on $\II$ on the arguments
$R^2\BB$ and $\hh_\infty\equiv h_\infty\LL/\HH_f$ ensures
that these relations constitute a pair of implicit algebraic
equations for $R=\RR_f/\LL$ and $\HH_f/\LL$. Moreover, 
the relations in (\ref{A6}) indicate that the functional
dependence on the parameters of the problem is through the
combinations $\BB\JJ^{-2/7}$ and $h_\infty\JJ^{-2/7}$ (or $h_\infty/\BB$),
as in (\ref{impo}).
Alternatively, one can map out the two functions on the
$(R^2\BB,\hh_\infty)$ parameter plane, and then interpolate
onto a grid of the physical parameters,
$\BB\JJ^{-2/7}$ and $h_\infty/\BB$, as done in figure \ref{daplot}.


\subsection{Edge structure}

\begin{figure}
    \centering
    \includegraphics[width=0.5\textwidth]{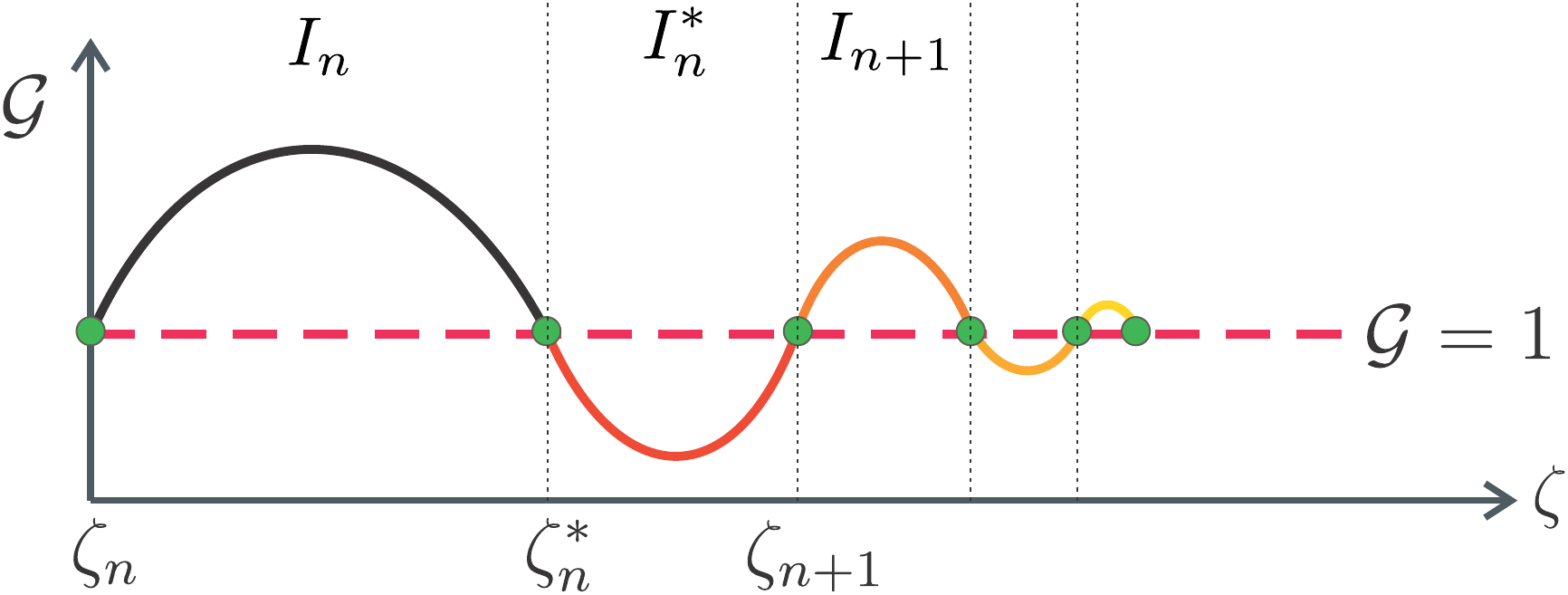}
    \caption{
      A sketch of the decaying undulations at the edge.
    }
    \label{fgr:VPoscilAsym}
\end{figure}

To analyse the structure at the edge, we consider the limit $h_\infty\ll1$
and resolve the narrow undulations there by introducing the
new variables $\mathcal{G}(\zeta)=h/h_{\infty}$
and $\zeta=(r-R)(\JJ/h_{\infty}^{2})^{1/3}$
For finite $\BB$ and to leading order, we then find
\begin{equation}
  \mathcal{G}_{\zeta \zeta \zeta} = \frac{1}{\mathcal{G}} \;
  \mathrm{sgn}(\mathcal{G}-1).
\label{vposcil}
\end{equation}
For $\mathcal{G} \rightarrow1$, (\ref{vposcil}) further simplifies to 
\begin{equation}
\mathcal{G}_{\zeta \zeta \zeta} = \mathrm{sgn}(\mathcal{G}-1).
\label{vposcil2}
\end{equation}
The wavetrain therefore limits to a sequence of cubic polynomials,
patched together to make the solution and its first two
derivatives continuous, as illustrated in figure \ref{fgr:VPoscilAsym}.
The approach to the pre-wetted film $\mathcal{G}=1$ is thereby achieved
by passing through an infinite sequence of switches in the
sign of $\mathcal{G}-1$, with
the second derivative $\mathcal{G}_{\zeta \zeta}$ taking a
decaying sawtooth waveform.

We split up the wave train into the intervals, $\{I_n\}$ and $\{I_n^*\}$,
between the sign switches of $\mathcal{G}-1$. 
The $n^{th}$ interval over which $\mathcal{G}>1$
is $I_n=[\zeta_n,\zeta_n^*]$, and the following one,
where  $\mathcal{G}<1$, is $I_n^*=[\zeta_n^*,\zeta_{n+1}]$.
We then write
\begin{equation}
\begin{split}
  I_{n}:  \mathcal{G} =
  1+\mathcal{G}^{\prime}_{n}(\zeta-\zeta_{n})
  + \frac{1}{2} \mathcal{G}^{\prime \prime}_{n}(\zeta-\zeta_{n})^{2}
  +\frac{1}{6} (\zeta-\zeta_{n})^{3},
  \\
  I_{n}^* : \mathcal{G} =
  1+\mathcal{G}^{\prime}_{n+1}(\zeta-\zeta_{n+1})
  + \frac{1}{2} \mathcal{G}^{\prime \prime}_{n+1}(\zeta-\zeta_{n+1})^{2}
  -\frac{1}{6} (\zeta-\zeta_{n+1})^{3},
\end{split}    
\end{equation} 
where primes denote the spatial derivatives with respect to $\zeta$
and the subscripts refer to the locations $\zeta_n$ where they
are evaluated. Matching these solutions together at $\zeta_{n}^{*}$
leads to a system of algebric equations:
\begin{equation}
\begin{split}
\mathcal{G}^{\prime}_{n}+\frac{1}{2} \mathcal{G}^{\prime \prime}_{n} z_{A} + \frac{1}{6} z_{A}^{2} = 0, \\
\mathcal{G}^{\prime}_{n+1}+\frac{1}{2} \mathcal{G}^{\prime \prime}_{n+1} z_{B} - \frac{1}{6} z_{B}^{2} = 0, \\
\mathcal{G}^{\prime}_{n}+ \mathcal{G}^{\prime \prime}_{n} z_{A} + \frac{1}{2} z_{A}^{2} - \mathcal{G}^{\prime}_{n+1}- \mathcal{G}^{\prime \prime}_{n+1} z_{B} + \frac{1}{2} z_{B}^{2}=0 \\
\mathcal{G}^{\prime \prime}_{n} + z_{A} + z_{B} = 0,\\
z_{A} = \zeta^{*}_{n} - \zeta_{n}, \quad z_{B} = \zeta_{n+1} - \zeta^{*}_{n}.
\end{split}
\label{algebsys}
\end{equation}
For large $n$, we seek the solution in the form,
\begin{equation}
\begin{split}
\mathcal{G}_{n}^{\prime} \sim \varrho^{2n} a \beta^{2}, \quad \mathcal{G}_{n}^{\prime \prime} \sim \varrho^{n} \beta \, \quad z_{A} \sim \varrho^{n} \beta b,\\ 
\mathcal{G}_{n+1}^{\prime} \sim \varrho^{2n+2} a \beta^{2}, \quad \mathcal{G}_{n+1}^{\prime \prime} \sim \varrho^{n+1} \beta ,\quad z_{B} \sim \varrho^{n+1} \beta c.
\end{split}
\label{algebsol}
\end{equation}
where $\beta$ is an arbitrary amplitude that must be fixed
by matching to the solution at the beginning
of the wavetrain. The remaining constants satisfy
\begin{equation}
\begin{split}
a+\frac{1}{2}b+\frac{1}{6}b^{2}=0,\quad
a+\frac{1}{2}c-\frac{1}{6}c^{2}=0,\\
a+b+\frac{1}{2}b^2-a\varrho^{2}-cr^{2}+\frac{1}{2}\varrho^{2}c^2=0,\quad
1+b-\varrho+\varrho \;c=0.
\end{split}
\end{equation}
These equations have a single set of valid solutions with $\varrho < 1$
(so that the undulations do not diverge), with
\begin{equation}
  \varrho = \frac{7}{2} -	\frac{3}{2}\sqrt{5} \approx 0.14,
  \quad a = 0.312, \quad b =-1.38, \quad c =3.618.
\end{equation}
Moreover, the distance along the wavetrain is given by
\begin{equation}
\zeta_{n} \sim \beta ( b  + c \varrho) \sum_{m=0}^{n} \varrho^{m}. 
\label{accEq}
\end{equation}
This geometric series has the finite limit,
\begin{equation}
  \underset{n \rightarrow \infty}{\rm Lim} \zeta_{n} =
  \frac{\beta ( b  + c \varrho)}{1-\varrho} = -\beta,
\end{equation}
implying the wavetrain ends at finite radius.

The preceding analysis establishes that the solution for the wavetrain
has a bounded solution in terms of the rescaled variables
$\zeta$ and $\mathcal{G}$. In the limit $h_\infty\to0$, the wavetrain
therefore shrinks to a point. Moreover,
given $(h,h_r,h_{rr})=(h_\infty\mathcal{G},\JJ^{1/3}h_\infty^{1/3}\mathcal{G}_\zeta,
\JJ^{2/3}h_\infty^{-1/3}\mathcal{G}_{\zeta\zeta})$,
the solution for $r<R$ must approach the edge with
$(h,h_r)\to0$ for $r\to R$,
but a diverging second derivative. Analysis of the singular point
of (\ref{eq:finalshapeEq}) at $r=R$ then establishes the local
solution used in \S \ref{3.2.2} to compute the solution
for $h_\infty\to0$.

\section{Gerris simulations}
\label{app:VOF}

\begin{figure}
    \centering
    \includegraphics[width=0.9\textwidth]{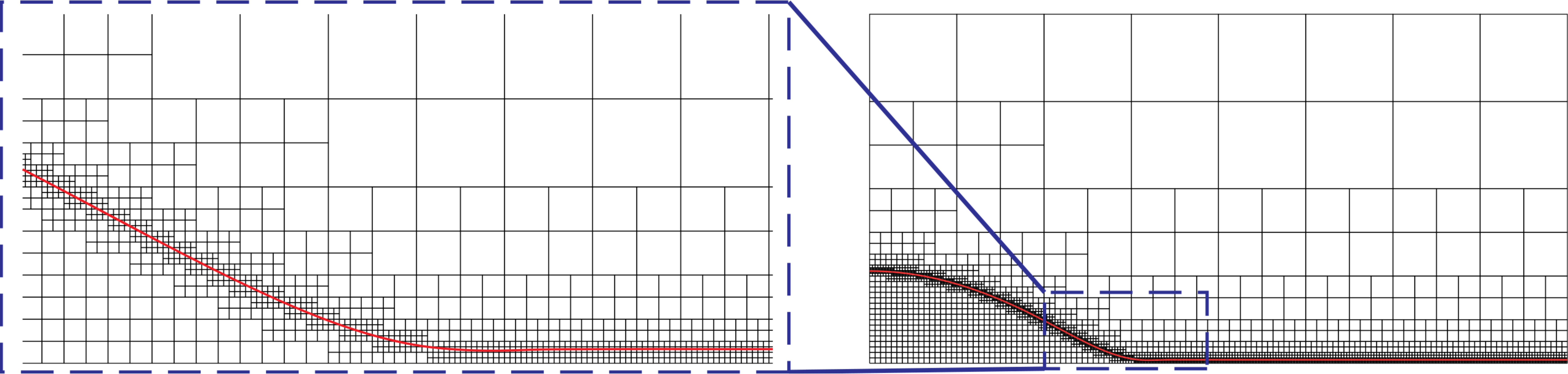}
    \caption{An example of numerical grids, where the maximum level in the grid generation was 8 over the interface of the droplet. Inside the droplet, the level of grids was always larger than 6. The interface corresponds to the case of $\mathcal{J}=0.144$ and $\mathcal{B}=0$ at $t=1.$ The magnified view of the grids around the edge of the droplet is shown in the right.}
    \label{fgr:gerrisGrid}
\end{figure}

The details of the Gerris simulations can be found in \citep{jalaal2016controlled}.
We use
rectangular domain that is sufficiently large to eliminate the influence
of the outer radial and top boundaries; see figure \ref{fgr:gerrisGrid},
which illustrates the adaptive gridding for an evolving solution.
Symmetry and no-slip boundary conditions were applied along the centreline and
bottom surface (respectively). ``Outflow'' conditions were applied on
the other two boundaries. We set the density and viscosity
of the ambient medium above the viscoplastic fluid
to be $10^{-2}\rho$ and $10^{-2}\mu$, respectively,
which ensures that this fluid does not influence the dynamics (as confirmed 
through a number of other simulations, changing the density and viscosity
ratios from $0.1$ to $0.002$). The computations are continued until
the value of kinetic energy fell below $10^{-6} \, \sigma\, \mathcal{L}^2$, at which point
a quasi-steady shape was largely established
and before any residual spreading arose due to the regularized viscosity
$\mu_{max}$.

\begin{figure}
    \centering
    \includegraphics[width=340pt]{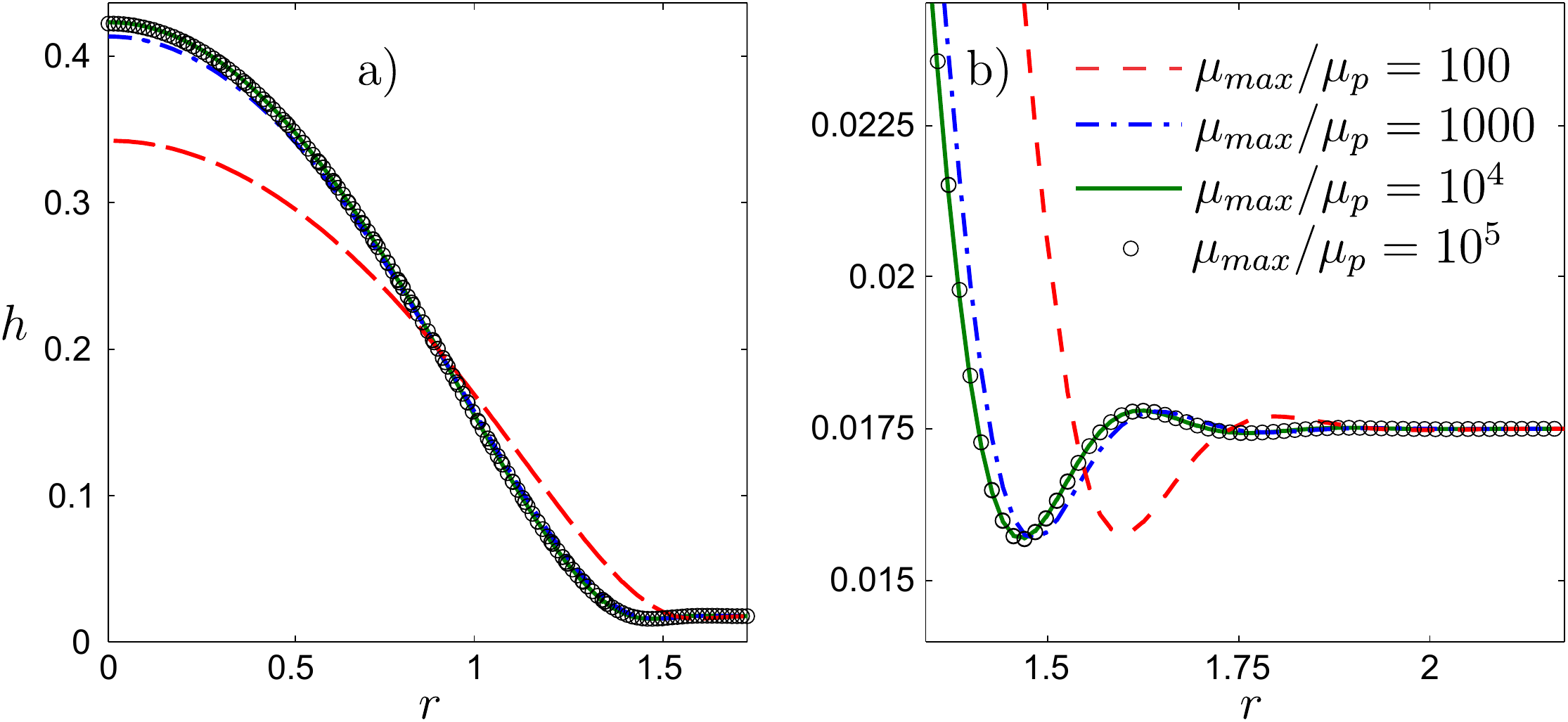}
    \caption{Effect of regularization parameter on the Gerris simulations.
      Results correspond to $\mathcal{J}=0.144$, $\mathcal{B}=0$ and
      $h_\infty=0.0175$.}
    \label{fgr:GerrisRegul}
\end{figure}

To verify that the simulations were independent of the
regularization parameter, we conducted numerical simulations for different
$\mu_{max}$ and established that the change to the final shapes were
insignificant for $\mu_{max} / \mu > 10^{4}$.
Figure \ref{fgr:GerrisRegul} shows an example of final shapes for different
regularization parameter.

\bibliographystyle{jfm}
\bibliography{refs}

\begin{thebibliography}{52}
\expandafter\ifx\csname natexlab\endcsname\relax\def\natexlab#1{#1}\fi
\def\au#1{#1} \def\ed#1{#1} \def\yr#1{#1}\def\at#1{#1}\def\jt#1{\textit{#1}}
  \def\bt#1{#1}\def\bvol#1{\textbf{#1}} \def\vol#1{#1} \def\pg#1{#1}
  \def\publ#1{#1}\def\arxiv#1{#1}\def\org#1{#1}\def\st#1{\textit{#1}}

\bibitem[Balmforth \& Craster(1999)]{balmforth1999consistent}
{\sc \au{Balmforth, NJ} \& \au{Craster, RV}} \yr{1999}  \at{A consistent
  thin-layer theory for bingham plastics}.  \jt{Journal of non-newtonian fluid
  mechanics}  \bvol{84}~(1),  \pg{65--81}.

\bibitem[Balmforth {\em et~al.\/}(2007{\natexlab{{\em a\/}}})Balmforth,
  Craster, Perona, Rust \& Sassi]{balmforth2007viscoplastic1}
{\sc \au{Balmforth, NJ}, \au{Craster, RV}, \au{Perona, P}, \au{Rust, AC} \&
  \au{Sassi, R}} \yr{2007{\natexlab{{\em a\/}}}}  \at{Viscoplastic dam breaks
  and the bostwick consistometer}.  \jt{Journal of non-newtonian fluid
  mechanics}  \bvol{142}~(1-3),  \pg{63--78}.

\bibitem[Balmforth {\em et~al.\/}(2006)Balmforth, Craster, Rust \&
  Sassi]{revslump}
{\sc \au{Balmforth, NJ}, \au{Craster, RV}, \au{Rust, AC} \& \au{Sassi, R}}
  \yr{2006}  \at{Viscoplastic flow over an inclined surface}.  \jt{Journal of
  non-newtonian fluid mechanics}  \bvol{139}~(1),  \pg{103--127}.

\bibitem[Balmforth {\em et~al.\/}(2014)Balmforth, Frigaard \&
  Ovarlez]{balmforth2014yielding}
{\sc \au{Balmforth, NJ}, \au{Frigaard, IA} \& \au{Ovarlez, G}} \yr{2014}
  \at{Yielding to stress: recent developments in viscoplastic fluid mechanics}.
   \jt{Annual Review of Fluid Mechanics}  \bvol{46},  \pg{121--146}.

\bibitem[Balmforth {\em et~al.\/}(2007{\natexlab{{\em b\/}}})Balmforth, Ghadge
  \& Myers]{balmforth2007surface}
{\sc \au{Balmforth, NJ}, \au{Ghadge, S} \& \au{Myers, T}}
  \yr{2007{\natexlab{{\em b\/}}}}  \at{Surface tension driven fingering of a
  viscoplastic film}.  \jt{Journal of non-newtonian fluid mechanics}
  \bvol{142}~(1),  \pg{143--149}.

\bibitem[Barnes(1999)]{barnes1999yield}
{\sc \au{Barnes, H~A}} \yr{1999}  \at{The yield stress—a review or
  ‘$\pi$$\alpha$$\nu$$\tau$$\alpha$
  $\rho$$\varepsilon$$\iota$’—everything flows?}  \jt{Journal of
  Non-Newtonian Fluid Mechanics}  \bvol{81}~(1),  \pg{133--178}.

\bibitem[Bergemann {\em et~al.\/}(2018)Bergemann, Juel \&
  Heil]{bergemann2018viscous}
{\sc \au{Bergemann, N}, \au{Juel, A} \& \au{Heil, M}} \yr{2018}  \at{Viscous
  drops on a layer of the same fluid: from sinking, wedging and spreading to
  their long-time evolution}.  \jt{Journal of Fluid Mechanics}  \bvol{843},
  \pg{1--28}.

\bibitem[Blackwell {\em et~al.\/}(2015)Blackwell, Deetjen, Gaudio \&
  Ewoldt]{blackwell2015sticking}
{\sc \au{Blackwell, BC}, \au{Deetjen, M~E}, \au{Gaudio, J~E} \& \au{Ewoldt,
  R~H}} \yr{2015}  \at{Sticking and splashing in yield-stress fluid drop
  impacts on coated surfaces}.  \jt{Physics of Fluids}  \bvol{27}~(4),
  \pg{043101}.

\bibitem[Blake(1990)]{blake1990viscoplastic}
{\sc \au{Blake, S}} \yr{1990}  \at{Viscoplastic models of lava domes}.  \bt{In
  {\em Lava flows and domes\/}},  \pg{pp. 88--126}.  \publ{Springer}.

\bibitem[Bonn {\em et~al.\/}(2017)Bonn, Denn, Berthier, Divoux \&
  Manneville]{bonn2017}
{\sc \au{Bonn, D}, \au{Denn, MM}, \au{Berthier, L}, \au{Divoux, T} \&
  \au{Manneville, S}} \yr{2017}  \at{Yield stress materials in soft condensed
  matter}.  \jt{Reviews of Modern Physics}  \bvol{89}~(3),  \pg{035005}.

\bibitem[Bonn {\em et~al.\/}(2009)Bonn, Eggers, Indekeu, Meunier \&
  Rolley]{bonn2009wetting}
{\sc \au{Bonn, Daniel}, \au{Eggers, Jens}, \au{Indekeu, Joseph}, \au{Meunier,
  Jacques} \& \au{Rolley, Etienne}} \yr{2009}  \at{Wetting and spreading}.
  \jt{Reviews of modern physics}  \bvol{81}~(2),  \pg{739}.

\bibitem[Boujlel \& Coussot(2013)]{boujlel2013}
{\sc \au{Boujlel, J} \& \au{Coussot, P}} \yr{2013}  \at{Measuring the surface
  tension of yield stress fluids}.  \jt{Soft Matter}  \bvol{9}~(25),
  \pg{5898--5908}.

\bibitem[Chen \& Bertola(2017)]{chen2017morphology}
{\sc \au{Chen, S} \& \au{Bertola, V}} \yr{2017}  \at{Morphology of viscoplastic
  drop impact on viscoplastic surfaces}.  \jt{Soft Matter}  \bvol{13}~(4),
  \pg{711--719}.

\bibitem[Coussot(2014)]{coussot}
{\sc \au{Coussot, P}} \yr{2014}  \at{Yield stress fluid flows: A review of
  experimental data}.  \jt{Journal of Non-Newtonian Fluid Mechanics}
  \bvol{211},  \pg{31--49}.

\bibitem[Coussot {\em et~al.\/}(2002)Coussot, Nguyen, Huynh \&
  Bonn]{coussot2002avalanche}
{\sc \au{Coussot, Philippe}, \au{Nguyen, Quoc~Dzuy}, \au{Huynh, HT} \&
  \au{Bonn, Daniel}} \yr{2002}  \at{Avalanche behavior in yield stress fluids}.
   \jt{Physical review letters}  \bvol{88}~(17),  \pg{175501}.

\bibitem[Craster \& Matar(2009)]{craster2009dynamics}
{\sc \au{Craster, RV} \& \au{Matar, OK}} \yr{2009}  \at{Dynamics and stability
  of thin liquid films}.  \jt{Reviews of modern physics}  \bvol{81}~(3),
  \pg{1131}.

\bibitem[Derby(2010)]{derby2010inkjet}
{\sc \au{Derby, B}} \yr{2010}  \at{Inkjet printing of functional and structural
  materials: fluid property requirements, feature stability, and resolution}.
  \jt{Annual Review of Materials Research}  \bvol{40},  \pg{395--414}.

\bibitem[Dimitriou \& McKinley(2019)]{dimitriou2019canonical}
{\sc \au{Dimitriou, C~J} \& \au{McKinley, G~H}} \yr{2019}  \at{A canonical
  framework for modeling elasto-viscoplasticity in complex fluids}.
  \jt{Journal of Non-Newtonian Fluid Mechanics}  \bvol{265},  \pg{116--132}.

\bibitem[Dinkgreve {\em et~al.\/}(2016)Dinkgreve, Paredes, Denn \&
  Bonn]{dinkgreve2016different}
{\sc \au{Dinkgreve, Maureen}, \au{Paredes, Jos{\'e}}, \au{Denn, Morton~M} \&
  \au{Bonn, Daniel}} \yr{2016}  \at{On different ways of measuring “the”
  yield stress}.  \jt{Journal of non-Newtonian fluid mechanics}  \bvol{238},
  \pg{233--241}.

\bibitem[Fraggedakis {\em et~al.\/}(2016)Fraggedakis, Dimakopoulos \&
  Tsamopoulos]{fraggedakis2016yielding}
{\sc \au{Fraggedakis, D}, \au{Dimakopoulos, Y} \& \au{Tsamopoulos, J}}
  \yr{2016}  \at{Yielding the yield stress analysis: A thorough comparison of
  recently proposed elasto-visco-plastic (evp) fluid models}.  \jt{Journal of
  Non-Newtonian Fluid Mechanics}  \bvol{236},  \pg{104--122}.

\bibitem[Geraud {\em et~al.\/}(2013)Geraud, Bocquet \&
  Barentin]{geraud2013confined}
{\sc \au{Geraud, B}, \au{Bocquet, L} \& \au{Barentin, C}} \yr{2013}
  \at{Confined flows of a polymer microgel}.  \jt{The European Physical Journal
  E}  \bvol{36}~(3),  \pg{30}.

\bibitem[G{\'e}raud {\em et~al.\/}(2014)G{\'e}raud, J{\o}rgensen, Petit,
  Delano{\"e}-Ayari, Jop \& Barentin]{geraud2014}
{\sc \au{G{\'e}raud, B}, \au{J{\o}rgensen, L}, \au{Petit, L},
  \au{Delano{\"e}-Ayari, H}, \au{Jop, P} \& \au{Barentin, C}} \yr{2014}
  \at{Capillary rise of yield-stress fluids}.  \jt{EPL (Europhysics Letters)}
  \bvol{107}~(5),  \pg{58002}.

\bibitem[German \& Bertola(2009)]{german2009}
{\sc \au{German, G} \& \au{Bertola, V}} \yr{2009}  \at{Impact of shear-thinning
  and yield-stress drops on solid substrates}.  \jt{Journal of Physics:
  Condensed Matter}  \bvol{21}~(37),  \pg{375111}.

\bibitem[Jalaal(2016)]{jalaal2016controlled}
{\sc \au{Jalaal, M}} \yr{2016}  \at{Controlled spreading of complex droplets}.
  PhD thesis, University of British Columbia.

\bibitem[Jalaal \& Balmforth(2016)]{jalaal2016long}
{\sc \au{Jalaal, M} \& \au{Balmforth, NJ}} \yr{2016}  \at{Long bubbles in tubes
  filled with viscoplastic fluid}.  \jt{Journal of Non-Newtonian Fluid
  Mechanics}  \bvol{238},  \pg{100--106}.

\bibitem[Jalaal {\em et~al.\/}(2015)Jalaal, Balmforth \&
  Stoeber]{jalaal2015slip}
{\sc \au{Jalaal, M}, \au{Balmforth, N~J} \& \au{Stoeber, B}} \yr{2015}
  \at{Slip of spreading viscoplastic droplets}.  \jt{Langmuir}  \bvol{31}~(44),
   \pg{12071--12075}.

\bibitem[Jalaal {\em et~al.\/}(2019{\natexlab{{\em a\/}}})Jalaal, Kemper \&
  Lohse]{jalaal2019vpentry}
{\sc \au{Jalaal, Mr}, \au{Kemper, D} \& \au{Lohse, D}} \yr{2019{\natexlab{{\em
  a\/}}}}  \at{Viscoplastic water entry}.  \jt{Journal of fluid mechanics}
  \bvol{864},  \pg{596--613}.

\bibitem[Jalaal {\em et~al.\/}(2019{\natexlab{{\em b\/}}})Jalaal, Seyfert \&
  Snoeijer]{jalaal2019capillary}
{\sc \au{Jalaal, M}, \au{Seyfert, C} \& \au{Snoeijer, J~H}}
  \yr{2019{\natexlab{{\em b\/}}}}  \at{Capillary ripples in thin viscous
  films}.  \jt{Journal of fluid mechanics}  \bvol{880},  \pg{430--440}.

\bibitem[J{\o}rgensen {\em et~al.\/}(2015)J{\o}rgensen, Le~Merrer,
  Delano{\"e}-Ayari \& Barentin]{jorgensen2015yield}
{\sc \au{J{\o}rgensen, L}, \au{Le~Merrer, M}, \au{Delano{\"e}-Ayari, H} \&
  \au{Barentin, C}} \yr{2015}  \at{Yield stress and elasticity influence on
  surface tension measurements}.  \jt{Soft matter}  \bvol{11}~(25),
  \pg{5111--5121}.

\bibitem[King(2001{\natexlab{{\em a\/}}})]{king2001b}
{\sc \au{King, John~R}} \yr{2001{\natexlab{{\em a\/}}}} The spreading of
  power-law fluids.  \bt{In {\em IUTAM Symposium on Free Surface Flows\/}},
  \pg{pp. 153--160}. Springer.

\bibitem[King(2001{\natexlab{{\em b\/}}})]{king2001a}
{\sc \au{King, John~R}} \yr{2001{\natexlab{{\em b\/}}}} Thin-film flows and
  high-order degenerate parabolic equations.  \bt{In {\em IUTAM Symposium on
  Free Surface Flows\/}},  \pg{pp. 7--18}. Springer.

\bibitem[Liu \& Mei(1989)]{liu1989slow}
{\sc \au{Liu, K~F} \& \au{Mei, C~C}} \yr{1989}  \at{Slow spreading of a sheet
  of bingham fluid on an inclined plane}.  \jt{Journal of fluid mechanics}
  \bvol{207},  \pg{505--529}.

\bibitem[Liu {\em et~al.\/}(2018)Liu, Balmforth \&
  Hormozi]{liu2018axisymmetric}
{\sc \au{Liu, Y}, \au{Balmforth, NJ} \& \au{Hormozi, S}} \yr{2018}
  \at{Axisymmetric viscoplastic dambreaks and the slump test}.  \jt{Journal of
  Non-Newtonian Fluid Mechanics}  \bvol{258},  \pg{45--57}.

\bibitem[Liu {\em et~al.\/}(2016)Liu, Balmforth, Hormozi \& Hewitt]{liu2016two}
{\sc \au{Liu, Y}, \au{Balmforth, NJ}, \au{Hormozi, S} \& \au{Hewitt, DR}}
  \yr{2016}  \at{Two--dimensional viscoplastic dambreaks}.  \jt{Journal of
  Non-Newtonian Fluid Mechanics} .

\bibitem[Luu \& Forterre(2009)]{luu2009}
{\sc \au{Luu, L.-H.} \& \au{Forterre, Y}} \yr{2009}  \at{Drop impact of
  yield-stress fluids}.  \jt{Journal of Fluid Mechanics}  \bvol{632},
  \pg{301--327}.

\bibitem[Luu \& Forterre(2013)]{luu2013giant}
{\sc \au{Luu, L.-H.} \& \au{Forterre, Y.}} \yr{2013}  \at{Giant drag reduction
  in complex fluid drops on rough hydrophobic surfaces}.  \jt{Physical review
  letters}  \bvol{110}~(18),  \pg{184501}.

\bibitem[Mackay(2018)]{mackay2018importance}
{\sc \au{Mackay, M~E}} \yr{2018}  \at{The importance of rheological behavior in
  the additive manufacturing technique material extrusion}.  \jt{Journal of
  Rheology}  \bvol{62}~(6),  \pg{1549--1561}.

\bibitem[Manglik {\em et~al.\/}(2001)Manglik, Wasekar \&
  Zhang]{manglik2001dynamic}
{\sc \au{Manglik, R~M}, \au{Wasekar, V~M} \& \au{Zhang, J}} \yr{2001}
  \at{Dynamic and equilibrium surface tension of aqueous surfactant and
  polymeric solutions}.  \jt{Experimental thermal and fluid science}
  \bvol{25}~(1),  \pg{55--64}.

\bibitem[O'Donovan \& Tanner(1984)]{o1984numerical}
{\sc \au{O'Donovan, EJ} \& \au{Tanner, RI}} \yr{1984}  \at{Numerical study of
  the bingham squeeze film problem}.  \jt{Journal of Non-Newtonian Fluid
  Mechanics}  \bvol{15}~(1),  \pg{75--83}.

\bibitem[Oishi {\em et~al.\/}(2019{\natexlab{{\em a\/}}})Oishi, Thompson \&
  Martins]{oishi2019impact}
{\sc \au{Oishi, CM}, \au{Thompson, RL} \& \au{Martins, FP}}
  \yr{2019{\natexlab{{\em a\/}}}}  \at{Impact of capillary drops of complex
  fluids on a solid surface}.  \jt{Physics of Fluids}  \bvol{31}~(12),
  \pg{123109}.

\bibitem[Oishi {\em et~al.\/}(2019{\natexlab{{\em b\/}}})Oishi, Thompson \&
  Martins]{oishi2019normal}
{\sc \au{Oishi, C~M}, \au{Thompson, R~L} \& \au{Martins, F~P}}
  \yr{2019{\natexlab{{\em b\/}}}}  \at{Normal and oblique drop impact of yield
  stress fluids with thixotropic effects}.  \jt{Journal of Fluid Mechanics}
  \bvol{876},  \pg{642--679}.

\bibitem[Oron {\em et~al.\/}(1997)Oron, Davis \& Bankoff]{oron1997long}
{\sc \au{Oron, A}, \au{Davis, S~H} \& \au{Bankoff, S~G}} \yr{1997}
  \at{Long-scale evolution of thin liquid films}.  \jt{Reviews of modern
  physics}  \bvol{69}~(3),  \pg{931}.

\bibitem[Popinet(2003)]{popinet2003gerris}
{\sc \au{Popinet, S}} \yr{2003}  \at{Gerris: a tree-based adaptive solver for
  the incompressible euler equations in complex geometries}.  \jt{Journal of
  Computational Physics}  \bvol{190}~(2),  \pg{572--600}.

\bibitem[Putz {\em et~al.\/}(2009)Putz, Frigaard \&
  Martinez]{putz2009lubrication}
{\sc \au{Putz, A}, \au{Frigaard, IA} \& \au{Martinez, DM}} \yr{2009}  \at{On
  the lubrication paradox and the use of regularisation methods for lubrication
  flows}.  \jt{Journal of Non-Newtonian Fluid Mechanics}  \bvol{163}~(1),
  \pg{62--77}.

\bibitem[Rafa{\"\i} {\em et~al.\/}(2004)Rafa{\"\i}, Bonn \&
  Boudaoud]{rafai2004spreading}
{\sc \au{Rafa{\"\i}, Salima}, \au{Bonn, Daniel} \& \au{Boudaoud, Arezki}}
  \yr{2004}  \at{Spreading of non-newtonian fluids on hydrophilic surfaces}.
  \jt{Journal of Fluid Mechanics}  \bvol{513},  \pg{77}.

\bibitem[Roussel \& Coussot(2005)]{roussel2005fifty}
{\sc \au{Roussel, N} \& \au{Coussot, P}} \yr{2005}  \at{“fifty-cent
  rheometer” for yield stress measurements: From slump to spreading flow}.
  \jt{Journal of Rheology (1978-present)}  \bvol{49}~(3),  \pg{705--718}.

\bibitem[Sa{\"\i}di {\em et~al.\/}(2010)Sa{\"\i}di, Martin \&
  Magnin]{saidi2010influence}
{\sc \au{Sa{\"\i}di, A}, \au{Martin, C} \& \au{Magnin, A}} \yr{2010}
  \at{Influence of yield stress on the fluid droplet impact control}.
  \jt{Journal of Non-Newtonian Fluid Mechanics}  \bvol{165}~(11),
  \pg{596--606}.

\bibitem[Saramito(2007)]{saramito2007new}
{\sc \au{Saramito, P}} \yr{2007}  \at{A new constitutive equation for
  elastoviscoplastic fluid flows}.  \jt{Journal of Non-Newtonian Fluid
  Mechanics}  \bvol{145}~(1),  \pg{1--14}.

\bibitem[Sen {\em et~al.\/}(2020)Sen, Morales \& Ewoldt]{sen2020viscoplastic}
{\sc \au{Sen, S}, \au{Morales, A~G} \& \au{Ewoldt, R~H}} \yr{2020}
  \at{Viscoplastic drop impact on thin films}.  \jt{Journal of Fluid Mechanics}
   \bvol{891}.

\bibitem[Tanner(1979)]{tanner1979spreading}
{\sc \au{Tanner, LH}} \yr{1979}  \at{The spreading of silicone oil drops on
  horizontal surfaces}.  \jt{Journal of Physics D: Applied Physics}
  \bvol{12}~(9),  \pg{1473}.

\bibitem[Thompson {\em et~al.\/}(2014)Thompson, Tipton, Juel, Hazel \&
  Dowling]{thompson2014}
{\sc \au{Thompson, Alice~B}, \au{Tipton, Carl~R}, \au{Juel, Anne}, \au{Hazel,
  Andrew~L} \& \au{Dowling, Mark}} \yr{2014}  \at{Sequential deposition of
  overlapping droplets to form a liquid line}.  \jt{Journal of fluid mechanics}
   \bvol{761},  \pg{261--281}.

\bibitem[Tuck \& Schwartz(1990)]{tuck1990numerical}
{\sc \au{Tuck, EO} \& \au{Schwartz, LW}} \yr{1990}  \at{A numerical and
  asymptotic study of some third-order ordinary differential equations relevant
  to draining and coating flows}.  \jt{SIAM review}  \bvol{32}~(3),
  \pg{453--469}.

\end{thebibliography}

\end{document}